\documentclass[mathleft]{an}

\usepackage{graphicx}
\usepackage{bm}
\usepackage[lofdepth,lotdepth]{subfig}
\usepackage{times}
\usepackage{amsmath}
\usepackage{amssymb}
\usepackage{natbib}
\usepackage{color}
\bibpunct{(}{)}{;}{a}{}{,}
\graphicspath{{./fig/}{./png/}}

\newcommand{\itover}[2]{\,\hspace{-.15mm}#1{\!\hspace{.15mm}#2}}%verallgemeinern!!

\newcommand{\ithat}[1]{\itover{\hat}{#1}}

\newcommand{\EQ}{\begin{equation}}
\newcommand{\EE}{\end{equation}}
\newcommand{\EQA}{\begin{eqnarray}}
\newcommand{\EEA}{\end{eqnarray}}
\newcommand{\brac}[1]{\langle #1 \rangle}
\newcommand{\pd}{\partial}

\newcommand{\DIV}{\vec{\nabla} \cdot }

\newcommand{\ve}[1]{\boldsymbol{#1}}
\newcommand{\dotm}[1]{\dot{\overline{#1}}}
\newcommand{\mean}[1]{\overline{#1}}

\newcommand{\Urms}{U_{\rm rms}}

\newcommand{\kef}{k_{\rm f}}

\newcommand{\Co}{{\rm Co}}

\newcommand{\Tay}{{\rm Ta}}
\newcommand{\Ray}{{\rm Ra}}
\newcommand{\Pra}{{\rm Pr}}
\newcommand{\Rey}{{\rm Re}}
\newcommand{\Nu}{{\rm Nu}}

\newcommand{\rxx}{\meanR_{xx}}
\newcommand{\ryy}{\meanR_{yy}}
\newcommand{\rzz}{\meanR_{zz}}
\newcommand{\rxy}{\meanR_{xy}}
\newcommand{\rxz}{\meanR_{xz}}
\newcommand{\ryz}{\meanR_{yz}}
\newcommand{\rij}{\meanR_{ij}}
\newcommand{\riz}{\meanR_{iz}}
\newcommand{\rjk}{\meanR_{jk}}

\newcommand{\rik}{\meanR_{ik}}
\newcommand{\trxx}{\tilde{\meanR}_{xx}}
\newcommand{\tryy}{\tilde{\meanR}_{yy}}
\newcommand{\trzz}{\tilde{\meanR}_{zz}}
\newcommand{\trxy}{\tilde{\meanR}_{xy}}
\newcommand{\trxz}{\tilde{\meanR}_{xz}}
\newcommand{\tryz}{\tilde{\meanR}_{yz}}

\newcommand{\Omx}{\Omega_x}

\newcommand{\Omz}{\Omega_z}

\def\onethird{{\textstyle{1\over3}}}

\def\onehalf{{\textstyle{1\over2}}}

\newcommand{\Uvec}{\vec{U}}
\newcommand{\uvec}{\vec{u}}
\newcommand{\gvec}{\vec{g}}
\newcommand{\nab}{\bm\nabla}
\newcommand{\meanRR}{\boldsymbol{\cal R}}
\newcommand{\meanR}{{\cal R}}
\newcommand{\meanFF}{\boldsymbol{\cal F}}
\newcommand{\meanF}{{\cal F}}
\newcommand{\meanQ}{{\cal Q}}
\newcommand{\meanT}{{\cal T}}
\newcommand{\meanU}{\mean{U}}
\newcommand{\LamF}{\Lambda_{\cal F}}
\newcommand{\LamR}{\Lambda_{\cal R}}
\newcommand{\LamQ}{\Lambda_\meanQ}
\newcommand{\phm}{\phantom{-}}

\newcommand{\meancT}{{\cal T}}
\newcommand{\hatg}{\ithat{g}}
\newcommand{\hatOm}{\hat{\Omega}}
\newcommand{\OO}{\bm{\Omega}}
\newcommand{\hatOO}{\hat{\OO}}
\newcommand{\hatgg}{\ithat{\vec{g}}}
\newcommand{\aap}{Astron. Astrophys.}
\newcommand{\pre}{Physical Review E}
\newcommand{\physscr}{Physica Scripta}

\begin{document}

\authorrunning{Snellman et al.}
\titlerunning{Testing turbulent closure models with convection simulations}

   \title{Testing turbulent closure models with convection simulations}

   \author{J. E. Snellman
	  \inst{1},
          P. J. K\"apyl\"a
	  \inst{1,2,3},
          M. J. K\"apyl\"a
	  \inst{3,1,2},
          M. Rheinhardt
	  \inst{1,2},
          \and
          B. Dintrans
          \inst{4}
	  }

   \institute{Department of Physics, Gustaf H\"allstr\"omin katu 2a 
              (PO Box 64), FI-00014 University of Helsinki, Finland
         \and NORDITA, Roslagstullsbacken 23, SE-10691 Stockholm, Sweden
        \and Aalto University, Department of Information and Computer Science, 
PO Box 15400, FI-00076 Aalto, Finland
         \and Observatoire Midi-Pyr\'en\'ees, Laboratoire d'Astrophysique de Toulouse-Tarbes (UMR5572), 14 Avenue Edouard Belin, 31400 Toulouse, France}

\received{} \accepted{}

   \abstract{
We compare 
     simple analytical closure models of homogeneous turbulent
     Boussinesq convection for stellar applications with 
     three-dimensional simulations.
   We use simple analytical closure models to compute the fluxes of
     angular momentum and heat as a function of rotation
     rate
     measured by
     the Taylor number. 
     We also investigate cases with varying angles between the angular
     velocity and gravity vectors, corresponding to locating the
     computational domain at different latitudes ranging from the pole
     to the equator of the star.
     We perform three-dimensional numerical simulations in the same
     parameter regimes for comparison. The free parameters appearing
     in the closure models are calibrated
     by two fitting methods using simulation data.
      Unique determination of the closure parameters is
     possible only in the non-rotating case 
     or
     when the system is
     placed at the pole. 
     In the other cases the 
     fit
     procedures yield somewhat differing results. 
     The quality of the
     closure is tested by 
     substituting the resulting coefficients
     back into the closure
     model and comparing with the 
     simulation
     results.
     To eliminate the possibilities that the results obtained
     depend on the aspect ratio of the simulation domain or    
     suffer from  too small Rayleigh numbers
     we performed
     runs varying these parameters.
   The simulation data for the Reynolds stress and heat fluxes
     broadly agree with previous compressible
     simulations.
     The closure works fairly well 
     with slow 
     and fast
     rotation but its quality degrades 
     for intermediate
     rotation rates.
     We find that the closure parameters
     depend not only on 
     rotation rate
     but also on latitude. 
     The weak dependence on Rayleigh number and the aspect ratio of
     the domain indicates that our results are generally valid.
    }


   \maketitle

%________________________________________________________________

\section{Introduction}

Turbulent convection is responsible for the transport of angular
momentum and heat
in stellar convection zones, in particular in that of the Sun.
In combination with global rotation, these turbulent flows lead to the
generation of large-scale differential rotation and meridional 
circulation
\citep[e.g.][]{R89}, which on the other hand, play key roles in 
sustaining the
dynamo of the Sun \citep[e.g.][]{KR80,RH04}.

During the last decades, growing computational resources have allowed
direct and large-eddy 
numerical
simulations to reach a level of sophistication
where many aspects of the solar differential rotation and dynamo can
be captured self-consistently 
\citep[see,e.g.][]{MBT06,MT09,GCS10,KMB12,WKMB13}.
However,
these simulations are still computationally very expensive and cannot
be 
employed
in performing comprehensive parameter surveys. Furthermore,
even the currently highest resolution simulations are still far
from real stars
in parameter space
 \citep[e.g.][]{K11}.  An alternate way of dealing with
the problem is to parametrize the small scales by turbulent transport
coefficients and solve 
directly
only for the large scales. 
This is often done by approximating higher-order correlations by lower order
ones in so-called {\em closure} approaches.
In general the pitfall lies in the limited
validity of the analytical approximations under which the
results are derived. Hence finetuning of the model
parameters is usually required.

There have been many different closure models proposed and used
through the years in astrophysical convection studies
\citep[e.g.][]{X1989,CM96,C1997,C2011,GOMS10}. These models differ
from each other by the approximations used (e.g.\ whether the fluid is
considered incompressible, anelastic, or fully compressible), and 
for which terms 
and in what way
closure approximations are invoked.
For instance, the closure studied in this paper \citep{GOMS10}, is based on
deriving exact time evolution equations for second-order correlations and
replacing the third-order ones 
occurring in those
by relaxation terms with a variable
relaxation time.

Another widely used model was introduced by \cite{CM96}, who started
from the evolution equations derived by \cite{Y1963} for the spectra
of the mean square velocity, mean square temperature perturbation and
turbulent heat flux in the Boussinesq approximation.  The nonlinear
transfer terms corresponding to the triple correlation terms $\langle
uuu \rangle $ and $\langle u\theta\theta \rangle $ (in symbolic notation)
were then equated to the spectra of mean square vorticity and mean
square temperature gradient, respectively, while the term of the type
$\langle uu\theta \rangle $ was assumed to depend on the two other
ones. For the closure ``parameters", a $k$-dependent turbulent viscosity
and heat conductivity were introduced.  These turbulent diffusivities
were considered to be related mutually by a single (constant) free
parameter.  Finally, the turbulent viscosity was set in relation to
the mean square vorticity with a coefficient derived from the
requirement that in the inertial range the resulting spectrum is to be
of Kolmogorov type.  In that way, the (normalized) turbulent heat
flux, rms value of the velocity, turbulent pressure etc. can be
determined explicitly after having chosen the free paramter.

A promising approach to the problem is to validate
or calibrate the turbulent closure models by comparing their results 
with direct numerical simulations. However, fairly little has been done to accomplish
this in the astrophysical context. This is especially true for
closures dealing with turbulent convection \citep[see, however,][]{GOMS10}.

In this paper we build upon previous studies where simple analytical
closure models were compared with 
simulations of forced or magnetorotationally excited turbulence in
fully periodic systems \citep{KB08,LKKBL09,SKKL09}. Here we extend
this work to turbulent convection in unstratified Rayleigh--B{\'e}nard
setups, drawing insights especially from the
results of
\cite{SBKM11,SRKMB12} (hereafter S12a and S12b) where closure
parameters were extracted from forced turbulence simulations. 
Our aim is to compare three-dimensional direct numerical simulations 
(DNS)
with the closure model for convection put forward by \cite{GOMS10}
(hereafter GOMS10). 
This
closure is an extension of earlier work
related to isothermal magnetohydrodynamic turbulence \citep{O03}. The
bulk 
of GOMS10
 is devoted to the
derivation and calibration of a closure model 
for a Boussinesq system. Results of DNS and experiments of bounded and
hence inhomogeneous non-rotating Rayleigh-B{\'e}nard convection are
referred to for the purpose of determining the free parameters of the
model. They found that in the statistically stationary
state, to a certain extent universal constants can be extracted which
moreover coincide partly
with those from a corresponding study of the very different situation of shear flows, see \cite{GO05}.

Similarly, an additional free parameter of the closure for {\em
  homogeneous} non-rotating Rayleigh--B{\'e}nard convection was
estimated on the
basis of DNS results where its universality turned out to be limited by the destabilization of the fluctuations in shallow computational domains.
The predictions of the closure model for the same setup, but with rotation included, were set into relation of previous analytical  results.
However, no direct comparison with corresponding DNS results was
performed, in particular, there was no independent calibration of the model
parameters.

The emergence of coherent structures covering the whole vertical extent of the domain was quite generally pointed out to be limiting the validity of 
the essentially {\em local} closure. For the case of rotating homogeneous Rayleigh-B{\'e}nard convection
a further  limit was found in the independence of the closure model  on the rotation rate when gravitation and rotation are perfectly aligned.

Hence, the goal of the present paper is to  scrutinize the potential of the GOMS10 closure model for rotating convection.

\section{Models and methods} \label{sec:model}
We consider a closure model for Boussinesq convection
in infinitely extended space
following 
\cite{MG07} and GOMS10. 

\subsection{The Boussinesq system} \label{sec:bouss}

\subsubsection{Basic equations}
In general,
the time evolution 
of the velocity and temperature fields is governed
by the Navier-Stokes, continuity, and heat transfer
equations
\begin{align}
\rho \frac{D \vec{U}}{D t} &= -\nab p +  \rho\,(\vec{g} - 2\, \bm{\Omega} \times \bm{U} + \vec{f}_{\rm visc}), \label{eq:bou1} \\
 \frac{D \ln \rho}{D t} &= - \nab \cdot \Uvec, \\ 
\rho c_V \frac{D T}{D t} &= \bm\nabla\cdot K \bm\nabla T -p \nab\cdot\vec{U}+ 2\nu \rho \bm{\mathsf{\bm S}}^2,\label{eq:temp}
\end{align}
where  $\vec{U}$
is the velocity, $\rho$ is the density,  
and
$\vec{g}$ is the gravitational
acceleration,
which is assumed constant. 
$\vec{f}_{\rm visc}$ is the viscous force per mass, $p$ is the
pressure, $T$ is the temperature, and $c_V$ is the specific heat at constant
volume, 
again assumed constant.
$K$ is the heat conductivity and
$D/Dt=\partial/\partial t+\vec{U}\cdot\bm{\nabla}$ denotes the
advective derivative. 
The viscous force is given by
\begin{equation}
\bm{f}_{\rm visc} = \nu \Big(\nab^2 \bm{U} +\onethird \vec{\nabla} \DIV \bm{U} + 2\, \bm{\mathsf{S}} \cdot \vec{\nabla} \ln \rho \Big),
\end{equation}
with the kinematic viscosity $\nu$,
assumed constant. $\bm{\mathsf{\bm S}}$ is the traceless rate of
strain tensor, which can be written in
component form as 
\begin{equation}
\mathsf{S}_{ij} = \onehalf \left(\pd_j U_i + \pd_i U_j \right) - \onethird \delta_{ij} \pd_k U_k\,.
\end{equation}

In the Boussinesq approximation, convection is understood as a perturbation to a stationary, purely conductive reference state
with  constant density $\rho_0$ which is hence governed by
\EQ
0= - \nab p_0 + \rho_0 \gvec, \quad 0= \chi \nab^2 T_0 + q_0, \label{eq:ref}
\EE
where the thermal diffusivity 
$\chi=K/\rho_0 c_V$
is assumed constant, and a stationary heat source $q_0$ can be included.
In this paper, however, we rely on the simplest case with $q_0=0$ and a consequently uniform temperature gradient $\nab T_0$ enforced by
appropriate boundary conditions. Denoting the deviations of density,
pressure and temperature from their reference values caused by
convection by $\rho'$, $p'$, and $\Theta$,
respectively, that is, $\rho=\rho_0+\rho'$, $p = p_0+p'$ and $T=T_0+\Theta$, we obtain from \eqref{eq:bou1} and \eqref{eq:ref}
for the momentum balance
\begin{align}
\rho\frac{D \vec{U}}{D t} &= - \nab p' + \rho' \vec{g} - 2\rho\bm{\Omega} \times \bm{U} + \rho\vec{f}_{\rm visc}.
\label{bou2}
\end{align}
According to the key idea of the Boussinesq 
approximation, the density deviation $\rho'$ from its reference value $\rho_0$ 
is assumed to be negligible {\em except in the buoyancy force} $ \rho \vec{g}$
\citep[see, e.g.,][]{Chandra61}.
Density and temperature perturbations are interconnected by
\begin{equation}
\frac{\varrho'}{\rho_0} = -\alpha(T-T_0) = -\alpha \Theta,
\label{bou3}
\end{equation}
where $\alpha$ is the coefficient of thermal
volume
expansion.
Finally, the equations of the Boussinesq approximation read
\begin{align}
\hspace*{-5mm}\frac{D \vec{U}}{D t} &= - \nab \Psi - \alpha \Theta \gvec - 2 \bm{\Omega} \times \bm{U} + \nu \nabla^2 \Uvec, \quad \nabla\cdot\Uvec=0,
\hspace*{-2mm}\label{eq:bouU} \\
\hspace*{-5mm}\frac{D \Theta}{D t} &= \chi \nab^2 \Theta - \Uvec\cdot\nab T_0 \, . \label{eq:bouThet}
\end{align}
Here, the reduced pressure $\Psi= p'/\rho_0$ was introduced and the viscous force was simplified for the now incompressible flow.

As first pointed out by \cite{SV60} this reasoning has to be modified
when being applied to gases rather than (practically incompressible)
liquids: While $\nab \cdot \Uvec=0$ is retained for the continuity
equation and the viscous force, the compression work
$-p\nab\cdot\vec{U}$ must not simply be omitted in \eqref{eq:bouThet}.
Instead, it gives rise to a twofold correction: For an ideal gas,
$\chi$ has to be redefined by employing $c_p$ instead of $c_V$ and the
background temperature gradient has to be replaced by the difference
$\nab T_0 - \vec{g}/c_p$ with the adiabatic temperature gradient
$\vec{g}/c_p$.  Thus we have
\EQ \frac{D \Theta}{D t} = \chi \nab^2\Theta - \Uvec\cdot\left(\nab T_0 - \frac{\vec{g}}{c_p} \right),
\label{eq:bou3}
\EE
with $\chi$ being now defined as $\chi = K/\rho_0 c_p $.

Contributions to the  heat budget from viscous heating are 
omitted in \eqref{eq:bouThet}, but can easily be taken in to account by restoring the term $2 \nu \bm{\mathsf{\bm S}}^2 /c_V$.
However, in order to guarantee energy conservation, expansion work has then also to be included in the form of  a cooling term  $\alpha \Theta \vec{g}\cdot\bm{U}$.

\subsubsection{Domain, boundary conditions, control parameters}
\label{sec:domain}
For the computational domain we consider a rectangular box thought of being
cut out at a varying latitude from the convection zone of a rotating star.
 We choose Cartesian coordinates  $(x,y,z)$ such that  their directions locally
 correspond to those of the global spherical coordinates $(\vartheta,\phi,r)$ having their axis $\vartheta=0$ aligned with the angular velocity vector $\bm{\Omega}$.
In the local Cartesian coordinates, the latter then reads
 \EQ
 \bm{\Omega} = \Omega_0(-\sin \vartheta,0,\cos \vartheta)^T,
 \label{OmegaDef}
 \EE
 where $\vartheta$ is the colatitude.  Gravity $\gvec$ is always
 taken to be radial, that is, to coincide with the local $z$
 direction.  We place the box at seven different positions defined by
 varying $\vartheta$ in equidistant steps of 15 degrees from $0 \degr$
 (pole) to $90\degr$ (equator).  For the dimensions of the box, $L_x$,
 $L_y$, $L_z$, we set $L_x=L_y$ while $L_z$ may be varied, see Section
 \ref{sec:aspectr}.  For all quantities, periodic boundary conditions
 in all directions are employed throughout the paper. If now the
 reference temperature gradient $\nab T_0$ is assumed constant over
 the box, that is, in the infinite space, this choice implies that the
 turbulence is {\em homogeneous} (but still anisotropic because
 vertical gravity introduces a preferred direction).  Hence this setup
 is labelled as {\em homogeneous Rayleigh-B{\'e}nard convection}.
We note that this type of system is not realizable in
   nature due to the periodic boundaries as discussed in
   \cite{CDGLTT06}. However,  due to its simplicity this setup is particularly useful in 
   testing closure models.

The system \eqref{eq:bouU}, \eqref{eq:bou3} is governed by the following three dimensionless parameters: The magnitude of the temperature 
gradient (and eventually the vigour of the convection) is quantified by the Rayleigh number
\EQ
\Ray = \frac{\alpha g d^3 (\Delta T_0-g d/c_p)}{\nu \chi},  
\label{eq:simray}
\EE
where $g=g_z$ and $d=L_z$ is the vertical extent of the domain.
In general, $\Delta T_0$ is the reference temperature difference
between its top and bottom.  For the homogeneous case considered here,
the definition \eqref{eq:simray} has to be modified properly employing
the constant effective (= prescribed minus adiabatic) background
temperature gradient $G_0=\nabla_z T_0-g/c_p$, that is
\EQ 
\Ray =\frac{\alpha g d^4 G_0}{\nu \chi}.
\label{eq:clmray}
\EE

The ratio of viscosity and thermal diffusivity is given by the Prandtl number
\EQ
{\rm Pr} =\frac{\nu}{\chi}\, , 
\EE
and finally the rotation rate is measured by the Taylor number
\EQ
{\rm Ta} =\frac{4\Omega^2 d^4}{\nu^2}.   \label{eq:Tay}
\EE
Another way to express the strengths of rotation and viscous effects,
but in the form of diagnostics rather than of control parameters,
is provided by the Coriolis and Reynolds numbers, respectively,
\EQ
 \Co = \frac{2\, \Omega_0}{\Urms \kef}\;, \quad
{\rm Re} = \frac{\Urms}{\nu \kef}\;,   
\label{eq:Corioliscomnew}
\EE
where $\kef=2\pi/d$ is the wave number corresponding to the vertical 
extent $d$ and $\Urms=(\int_V U^2 dV/V)^{1/2}$ is the 
root mean square velocity with the volume of the domain $V$.
The
efficiency
of the convective heat transfer
is measured by the Nusselt number
estimating the ratio of the total to the conductive heat flux
\EQ
{\rm Nu} =\frac{\brac{U_z\Theta}}{\chi G_0}+1,   \label{eq:Nu}
\EE
where the angle brackets denote volume averaging.

\subsubsection{Closure model} \label{sec:closure}

In this section we present the homogenous version of the GOMS10 model.
The details of the more general 
inhomogenous model can be found in Appendix \ref{app:eq}.
First we
specify the averaging procedure
by which the mean quantities are defined. 
Given the homogeneity of our model, volume averages are 
applied to the numerical results
throughout this paper. Hence, the mean of a quantity $f$, indicated by an overbar, is given by $\mean{f}=\int_{L_z}\! \int_{L_y} \!\int_{L_x} \! f(x,y,z) dx dy dz /L_x L_y L_z$.
This procedure satisfies all the Reynolds averaging rules. 
Denoting fluctuating quantities with lowercase letters, we have $\vec{U}=\mean{\vec{U}}+\vec{u}$,  $\Theta=\mean{\Theta}+\theta$ etc. 
When assuming $\Omega_y=0$,
and the temperature gradient is defined to be a constant vector parallel 
to gravity 
$\vec{g}=g \vec{e}_z$, hence
 $\nab T_0 -\gvec/c_p = G_0 \vec{e}_z$,
the equations for the 
mean velocity 
and temperature
resulting from \eqref{eq:bouU}, \eqref{eq:bou3}
read
\begin{alignat}{2}
\dotm{U}_x &= 2\Omz \mean{U}_y ,& \quad 
\dotm{U}_y &= - 2\Omz \mean{U}_x + 2\Omx \mean{U}_z, \nonumber \\
\dotm{U}_z &= -\alpha\mean\Theta g -2\Omx \mean{U}_y, &\quad 
\dotm{\Theta} &=  - \mean{U}_{\!z} G_0,  \hspace{-3mm} \label{eq:Uxy} 
\end{alignat} 

cf. also \eqref{eq:Uxy2}, \eqref{eq:thet2}.
As a consequence of the chosen average,
spatial derivatives vanish and the continuity equation is satisfied automatically.
The system \eqref{eq:Uxy} does not invoke
the Reynolds stress tensor $\rij=\mean{u_i u_j}$,
the turbulent heat flux, $\meanF_i=\mean{\theta u_i}$
or the temparature variance $\meanQ=\mean{\theta^2}$ and is therefore closed.
As it is homogeneous, its solutions vanish if the initial
conditions do so, but as it possesses unstable solutions, it is necessary to suppress them
explicitly in the DNS.

Evolution equations for the Reynolds stress and turbulent heat flux 
can be derived from the equations for the 
fluctuating
quantities $\uvec$ and 
$\theta$, see Appendices~\ref{app:eq}
and \ref{app:eq0}.
In doing so one comes inevitably across higher-order correlations of
$\uvec$ and $\theta$.
The essential step of the closure procedure as proposed in GOMS10
consists then in replacing these correlations by
aggregates of second-order correlations, more specifically, of the quantities 
$\meanR_{ij}$ and $\meanF_i$ themselves. In an analogous way, some second-order 
correlations which cannot directly be expressed by the components of $\meanRR=\{\meanR_{ij}\}$ 
and $\meanFF=\{\meanF_i\}$
are modelled. 
In order to obtain a closed system, an additional equation 
for
$\meanQ$
is needed which can be derived using Eq.~\eqref{eq:bou3}
and which is also subjected to the closure procedure.
The details can again be found in Appendix~\ref{app:eq}.
Finally,
the closed set of
equations reads
\begin{align}
\dot{\meanR}_{ij} &+   \alpha (\meanF_i g_j + \meanF_j g_i)  + 2 \Omega_{l}\! \left(\varepsilon_{ilk} \rjk + \varepsilon_{jlk} \rik \right)\label{eq:clor} \\ 
&=  -\left( \frac{C_1+C_2}{L} \meanR^{1/2} + \nu \frac{C_\nu}{L^2} \right)\rij + \frac{C_2}{3L} \meanR^{3/2} \delta_{ij} , \nonumber \\[3mm]
 \dot{\meanF}_i &+ \riz G_0 + \alpha \meanQ g_i + 2\varepsilon_{ijk}\Omega_j \meanF_k \label{eq:clof}\\ 
 &= - \left(\frac{C_6}{L} \meanR ^{1/2}  + \onehalf (\nu + \chi) \frac{C_{\nu\chi}}{L^2}\right) \meanF_i, \nonumber\\
\dot{\meanQ} &+ 2\meanF_z G_0  =  -\left(\frac{C_7}{L} \meanR^{1/2}  + \chi \frac{C_{\chi}}{L^2} \right)\!\meanQ , \label{eq:cloq}
\end{align}
where $\meanR =\rxx+\ryy+\rzz$ is the trace of $\meanRR$, 
$C_{1,2,6,7,\nu,\nu\chi,\chi}$ are the model parameters 
and $L$ is a characteristic length scale.
Note that this system does not invoke $\mean\Uvec$ or $\mean\Theta$ so their potentially unstable behavior is
irrelevant here.
In GOMS10 the 
model parameters
were assumed to be universal constants and $L$ was taken to be proportional to 
the shortest length scale of the simulation box, $L=\min(L_x, L_y, L_z)$.

Apart from the isotropic tensor $\delta_{ij}$, the closure terms in \eqref{eq:clor} contain only the tensor $\meanRR$ and thus 
do not explicitly reflect
anisotropies which
could be induced by 
$\meanFF$, or by preferred
directions present already in the setup like $\bm{\Omega}$ and
$\gvec$.
The same holds {\it mutatis mutandis} for the closure terms in \eqref{eq:clof}.
A complete formulation would need to be built up from quite a number of 
tensorial building blocks, each accompanied with a coefficient.
Possible terms up to second order in the unit vectors
$\hatOO=\OO/\Omega$, $\hatgg=\vec{g}/g$,
obeying
the constraint that
no other pseudoscalar
than $\cos\vartheta=\hatgg\cdot\hatOO$ is available, further without
cross-influences of mean quantities, i.e., without using
$\meanFF$ in the closure for $\meanRR$
and vice versa,
are the following for $\dot{\meanR}_{ij}$:
\begin{align}
&\hspace*{-1mm}\begin{alignedat}{2}
  & \hatg_i \hatg_j,\; \hatOm_i \hatOm_j &&\hspace*{-0mm} \text{with coefficients}\;\: \sim \meanR^{3/2}/L \\
  &  (\meanR_{il} \hatg_j+\meanR_{jl} \hatg_i)\hatg_l , \;  (\meanR_{il} &&\hspace*{-0mm} \hatOm_j+ \meanR_{jl} \hatOm_i) \hatOm_l, \\
  &   (\meanR_{il} \varepsilon_{lkj}  +\meanR_{jl} \varepsilon_{lki}) \hatOm_k 
  &&\hspace*{-0mm} \text{with coefficients}\;\: \sim \meanR^{1/2}/L
\end{alignedat}\hspace{-1cm} \label{eq:extR}\\[-1mm]
\intertext{and for $\dot{\meanF}_{i}$:}
&\begin{alignedat}{2}
  &\hatg_i , \; \varepsilon_{ikl}\hatg_k \hatOm_l &&\text{with coefficients}\;\: \sim \meanR^{1/2} \meanF/L \\
  &\hatg_j \meanF_j \hatg_i, \; \hatOm_j \meanF_j \hatOm_i,  \; &&\\
&\varepsilon_{ijk}\hatOm_j \meanF_k \;\;&&\text{with coefficients}\;\: \sim \meanR^{1/2}/L\,.
\end{alignedat}\label{eq:extF}
\end{align}
At the pole (where $\OO \!\parallel\! \vec{g}$) a
stationary solution of \mbox{\eqref{eq:clor}--\eqref{eq:cloq}} is given in \eqref{eq:specpole1}--\eqref{eq:specpole2}
where only the four quantites  $\rxx=\ryy$, $\rzz$, $\meanF_z$ and $\meanQ$ are different from zero. 
This solution agrees qualitatively with
corresponding 
 DNS results (see Sec.~\ref{sec:DNSruns}). 
Note that a closure, {\em extended} by the terms listed in \eqref{eq:extR}, \eqref{eq:extF}, would still 
 allow for such a
solution.
Several of the
terms listed in \eqref{eq:extR}, \eqref{eq:extF} vanish at the pole
and for all the remaining ones there are structurally identical terms
within the original closure. Consequently, the coefficients of
these additional terms, 
 can all be
absorbed in the $\{C_i\}$.

  Already in GOMS10, the 
  original
  closure 
  \eqref{eq:clor}--\eqref{eq:cloq}
  was
  found to miss the reduction of the convective heat flux in the
  presence of rotation at the pole.
  This is an inevitable consequence of the assumption of universal 
  closure parameters, as at this location rotation is not showing up explicitly
  in the closure equations.
  Hence some
  modification of the model is clearly necessary. In \cite{MG07} a
  corresponding attempt was undertaken by assuming the length scale
  $L$ of the model to be dependent on the wavelength of the most
  unstable convective eigenmode $\lambda$, which 
  in turn
  depends on rotation
  rate, thus making $L$ also a function of rotation rate, or
  $L=L(\Tay)$. Since all model coefficients $C_*$ appear in ratios
  $C_*/L$, a $\Tay$ dependence in $L$ can always be transferred to the
  $C_*$.  In \cite{MG07} only one of the possibilities for $L(\Tay)$
  was considered, namely a mean of $\lambda$ and the distance $d$ to
  the closest boundary, $L=\left( 1/d^2 + 2/\lambda^2 \right)$.

In this paper we 
adopt the view that all closure parameters must in general depend 
on all control parameters of the setup, namely $\Tay$, $\vartheta$, $\Ray$ and $\Pr$.
This is a natural lesson from mean field theory where coefficients,
parameterizing the turbulence, say, in the Reynolds stress,
are obtained from (approximate) solutions
for the fluctuating parts of the system quantities, that is in our context, $\uvec$ and $\theta$. 
Their governing equations
contain the control parameters as coefficients, hence the fluctuating parts are in general
dependent on them and thus also the mean-field coefficients, see \citep{M78,KR80,R89}.
As there is hardly a fundamental difference between mean-field coefficients and closure
parameters, the latter can neither be universal constants.

As a first step we retain the original structure of the closure \eqref{eq:clor}--\eqref{eq:cloq}, but allow its coefficients
$C_*$ to vary with the control parameters, calling this model ``minimally extended GOMS10 closure''.
We will in particular attempt to systematically identify
the Taylor number and latitude dependence 
of the 
$C_*$
and, to a lesser extent, also their dependence on $\Ray$.
As a way of extending the original GOMS10 closure to the rotating
case, this seems to be most straightforward and
rather easy to study,
and also
less restrictive than the approach of \cite{MG07}.

The onset of convection and its saturated stage as functions of rotation rate
were studied 
by these authors,
but they did not directly 
relate
mean quantities
like $\meanRR$
from DNS to the corresponding ones from the closure model.
Our approach is here, in contrast,
to derive the 
supposed control parameter dependences
of the closure parameters
referring directly to DNS results.

\begin{figure*}
\centering
\hspace*{-.7cm}\includegraphics[width=0.367\textwidth]{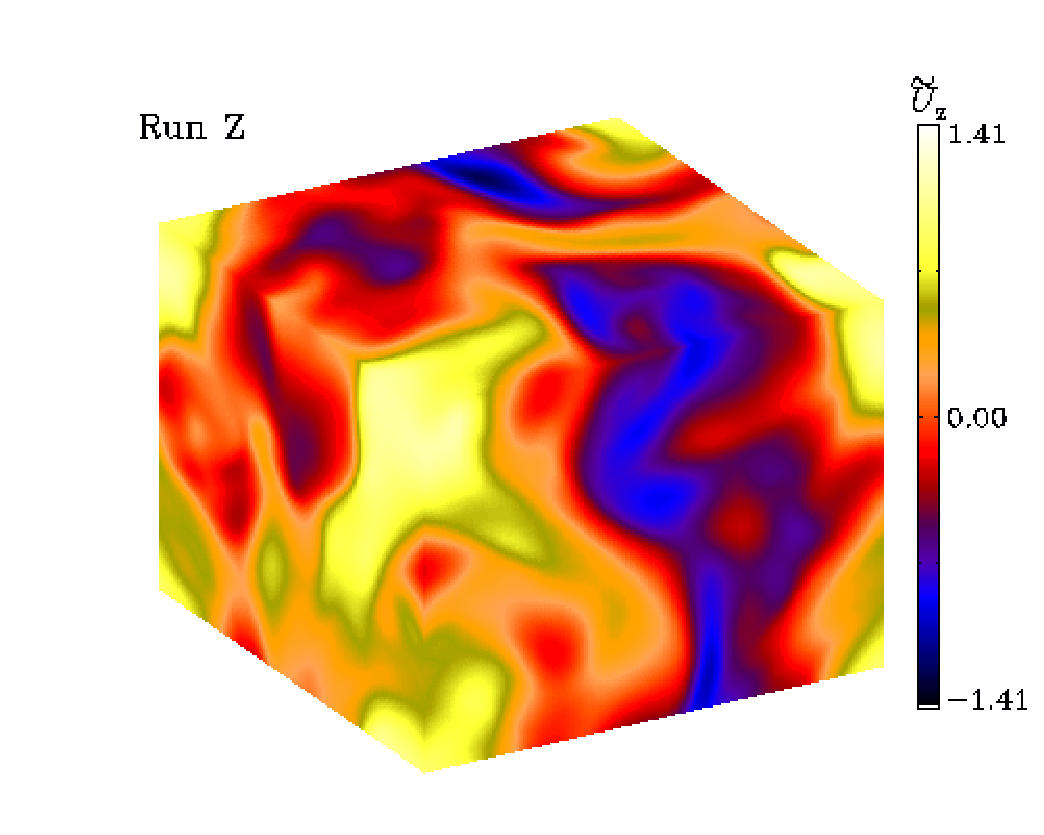}\includegraphics[width=0.367\textwidth]{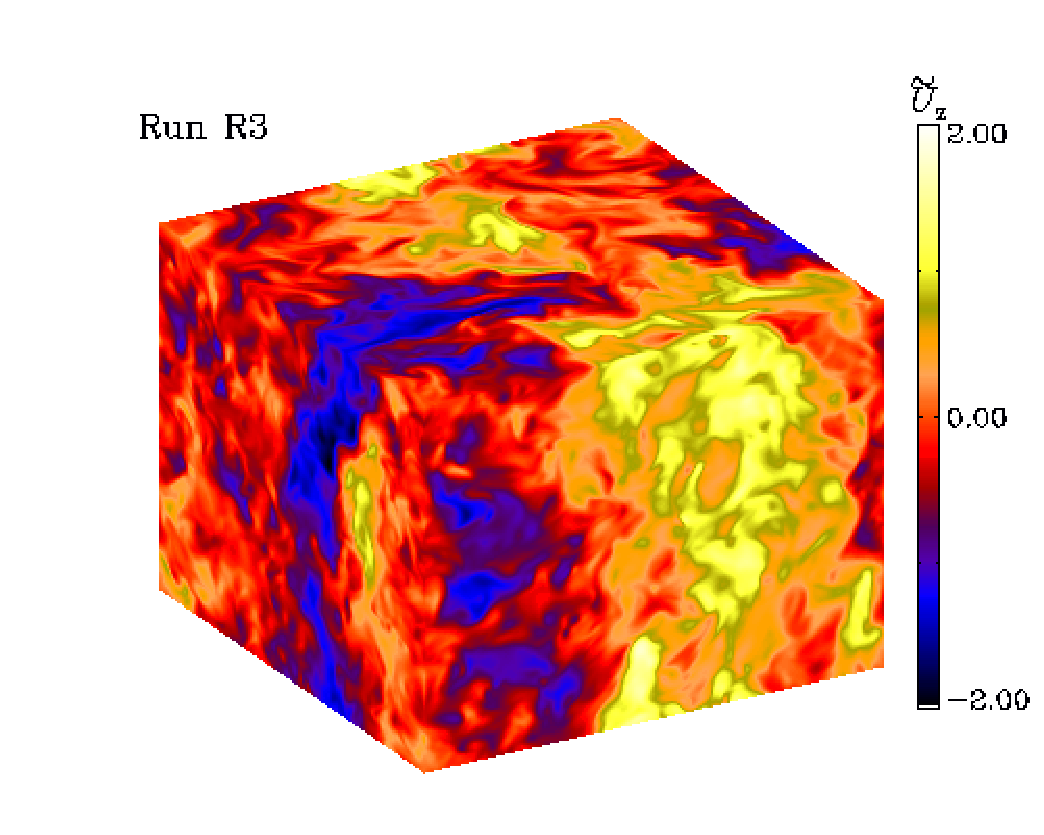}\includegraphics[width=0.367\textwidth]{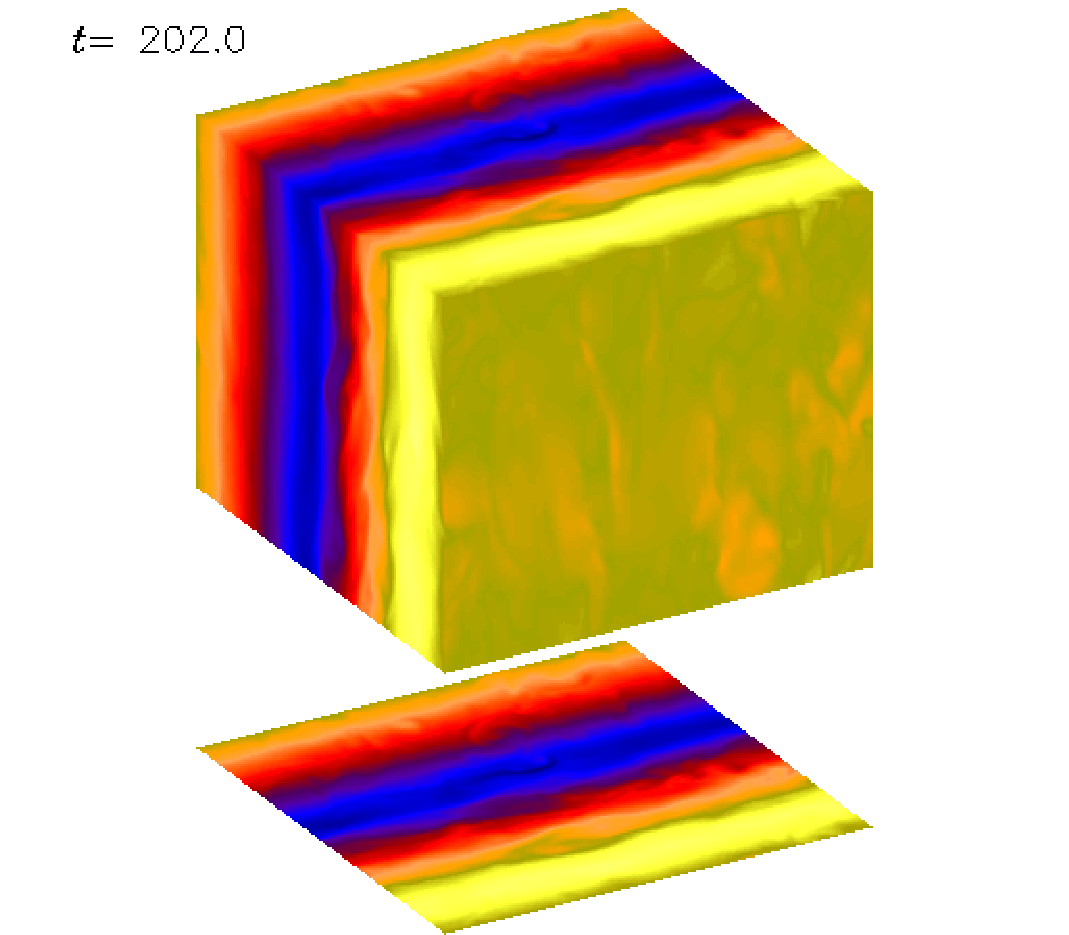}
\caption{Velocity component $U_z$ in units of $d(\alpha g  G_0)^{1/2}$ from non-rotating DNS runs Z (left) and R3 (center) 
  with $\Ray=3\cdot10^5$ and $\Ray=2.5\cdot10^7$, respectively. The figure on the right is from an experimental Run with very high rotation rate ($\Tay = 10^{10}$) at $\vartheta = 75^{\degr}$.}
\label{fig:uzb}
\end{figure*}

\subsection{DNS setups}
\label{sec:convmodel}

For the DNS
a local Cartesian
volume of size $L_x \times L_y \times L_z$
is used with $L_x = L_y$ and aspect ratio $\Gamma=L_z/L_x=1,4$
with
 fully periodic 
 boundary conditions and 
 a uniform background temperature gradient
as described in Sec.~\ref{sec:bouss}.
Grid sizes ranging from $64^3$ and $512^3$ are used,
  with the latter corresponding to runs with the highest Rayleigh
  numbers.

\begin{figure}
\centering
\includegraphics[width=\columnwidth]{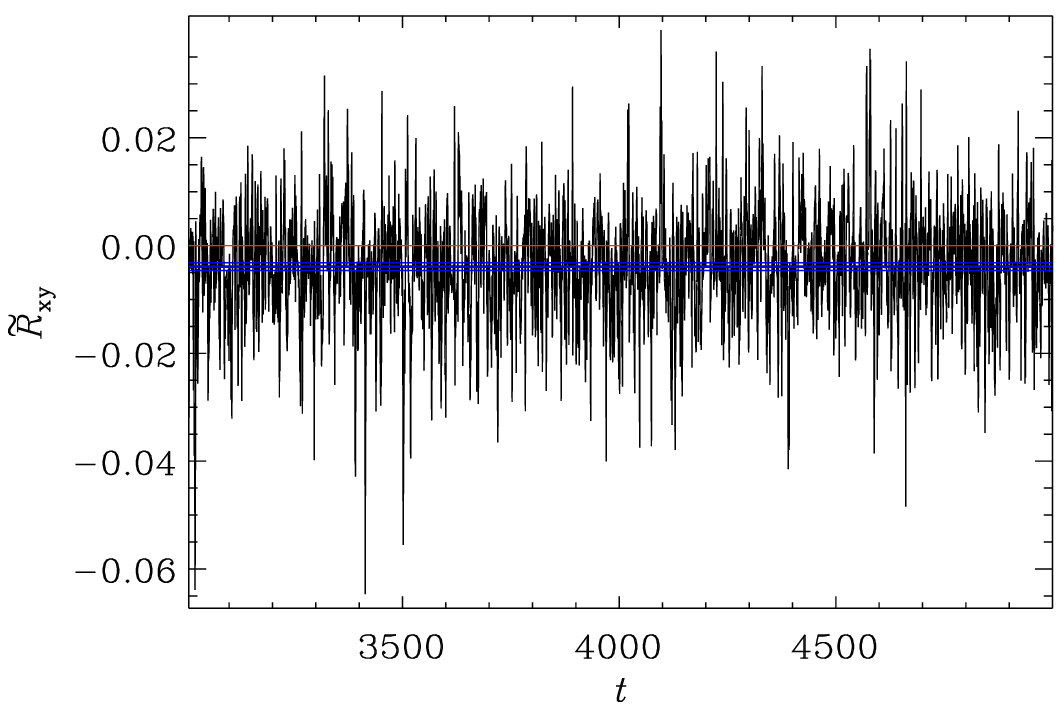}
\caption{Time series of $R_{xy}$ from Run~B4. The blue
  solid lines show the average and the error estimates whereas the red
  solid line denotes the zero level.}
\label{fig:perror} 
\end{figure}

The numerical simulations were performed with the {\sc Pencil
  Code}\footnote{http://code.google.com/p/pencil-code/},
which uses sixth-order accurate finite differences in space, and a
third-order accurate time-stepping sche\-me,
see
 \citet{BD02,B03}.
 Originally designed for (weakly) compressible hydrodynamics, it has recently
 been supplemented by a module implementing the Boussinesq approximation,
 following a method presented in \cite{BM92}.
In all cases 
the time integration was advanced until a statistically stationary state was reached.
Typically this means at least a few hundred convective turnover times. 
In addition to the 
volume averages, time averages over
 this state are taken
because 
the spatial averages still show strong fluctuations.
Errors are estimated by dividing the time series into three equally
long parts and computing mean values for each part individually. The
largest departure from the mean value computed for the whole time
series is taken to represent the error.
A representative example is shown in
  Figure~\ref{fig:perror}.

%________________________________________________________________
\section{Results}
\label{sec:results}

\subsection{DNS Runs}
\label{sec:DNSruns}

The DNS
runs are summarised in Table~\ref{tab:simsetsB},
for more details see Table~\ref{tab:bres} in the Online Material.
In each of the sets A--G the latitude was kept fixed, but the rotation
rate was varied. Z denotes the non-rotating run. The ranges for the
Reynolds, Rayleigh, and
Coriolis numbers as well as
the rms velocity $\Urms$
probed by DNS are also listed.

In the 
DNS runs we 
had to deal with a
numerical stability problem
in the transition from
the kinematic, exponentially growing, stage to the stationary stage if
the Taylor number was too small.
To circumvent this 
we started the runs 
with high ($\approx10^7$) values of $\Tay$ which allowed a
statistically stationary
state to be established. Then we gradually lowered $\Tay$ until the
desired parameter range had been reached.
Eventually we were 
able to successfully perform non-rotating runs using this
method. A snapshot of the vertical velocity in
this
Run~Z is
pictured in Fig.~\ref{fig:uzb}.
In the time series of the statistically stationary state,
we observe large fluctuations and intermittent exponential growth,
most likely
as a manifestation of the so-called ``elevator modes", 
described in \cite{CDGLTT06}
and being exponentially growing solutions of the {\em nonlinear} Boussinesq system \eqref{eq:bouU}, \eqref{eq:bou3}.
For a box aspect ratio of unity
they are known to be only weakly damped.
In our setup, these modes exist uninfluenced by rotation at any latitude in the form
When admitting a horizontal velocity component, 
another type of
exponentially growing solutions of the nonlinear equations 
is possible  at  the equator, having
the form 
$[0,U_y(x),U_z(x)]\!\sim\!\exp(\lambda t+{\rm i} k_x x)$.
The critical Rayleigh number is then given by $(dk_x)^4 + \Tay$. 
In any case the elevator modes should 
be affected
by rotation, insofar as the
secondary instabilities which are responsible for
their
ultimate saturation
will also certainly be modified by the Coriolis force.

The qualitative behaviour of the elevator modes can be demonstrated by
the rightmost panel of Fig.~\ref{fig:uzb} which shows a rapidly
rotating 
($\Tay = 10^{10}$) Run at $\vartheta = 75^{\degr}$.
The depicted large-scale flow pattern appears and disappears regularly
in our numerical simulations, coinciding with the large fluctuations
mentioned above. Curiously
enough, this pattern manifests itself ever more clearly when rotation rate is increased: At the lowest rotation rates, it is only seen blurred by the usual small fluctuations 
of the system, while in the intermediate rotation rates its periodical appearance and reconfiguration blots the timeseries with periods of exponential growth and decay. 
At very high rotation rates the stripe feature becomes permament, accompanied by very high values for velocity components and every other quantity measured from the Run.
This is why some runs in the rapid rotation regime needed to be
omitted, i.e.\ the flow pattern was completely dominated by the
elevator modes and no turbulence was present, see Table~\ref{tab:bres}
for the details.

The DNS were further complicated by convergence issues. Increasing
only the spatial resolution had only a minor effect in the results
even at largest values of $\Tay$. However, reducing time step $\delta
t$ affected the results more dramatically requiring many of the
simulations to run for far longer than initially expected. This
problem became more pronounced in the rapid rotation regime. We
obtained converged results by halving the time step until the results
from the two shortest $\delta t$ agreed within ten per cent.

%JES
\begin{table}
  \centering
  \caption[]{ \label{tab:simsetsB} Summary of the DNS runs at 
different colatitudes $\vartheta$. 
$\Pra=1$ for sets R and R', and $\Pra=0.6$ for all other sets.
$\Tay = 4\cdot10^4 \ldots 1.3\cdot10^7$ in sets A through G, $\Tay = 0$ for sets
Z and R, and $\Tay = 10^6$ for R'.
}
  \vspace{-0.5cm}
  $$
  \begin{array}{p{0.05\linewidth}ccccccccrrr@{\hspace{-10mm}}}
    \hline
    \hline\\[-2.5mm]
    \noalign{\smallskip}
     Set  &  \vartheta  & \Ray/10^{5}  & \Rey  & \Co &  \Urms/d\sqrt{\alpha g G_0\phantom{)}\!\!\!}\\
    \noalign{\smallskip}
    \hline\\[-2mm]
    Z  & -  & 3 & 91 & 0 &  0.57 \\
    A  &    \mbox{\phantom{3}0}\degr  & 3         & 91-127      & 0.06-0.87 & 0.57-0.80 \\
    \noalign{\smallskip}
    B  &    15\degr                   & 3         & 81-91       & 0.06-1.09 & 0.51-0.57 \\
    \noalign{\smallskip}
    C  &    30\degr                   & 3         & 70-89       & 0.06-1.07 & 0.44-0.56 \\
    \noalign{\smallskip}
    D  &    45\degr                   & 3         & 62-91       & 0.06-1.39 & 0.39-0.57 \\
    \noalign{\smallskip}
    E  &    60\degr                   & 3         & 57-96       & 0.06-0.95 & 0.36-0.61 \\
    \noalign{\smallskip}
    F  &    75\degr                   & 3         & 54-89       & 0.06-1.02 & 0.34-0.54 \\
    \noalign{\smallskip}
    G  &    90\degr                   & 3         & 48-85       & 0.06-1.22 & 0.30-0.54 \\
    \noalign{\smallskip}
    R  &  -                           & 2.5-2000  & 68-2358     & 0         & 0.80-1.04 \\
    \noalign{\smallskip}
    R' &  \mbox{\phantom{3}0}\degr    & 10-250    & 128-764     & 0.03-0.2  & 0.81-0.96 \\
    \hline
   \end{array}
   $$ 
\end{table}

\begin{figure*}
\centering
\includegraphics[width=.75\textwidth]{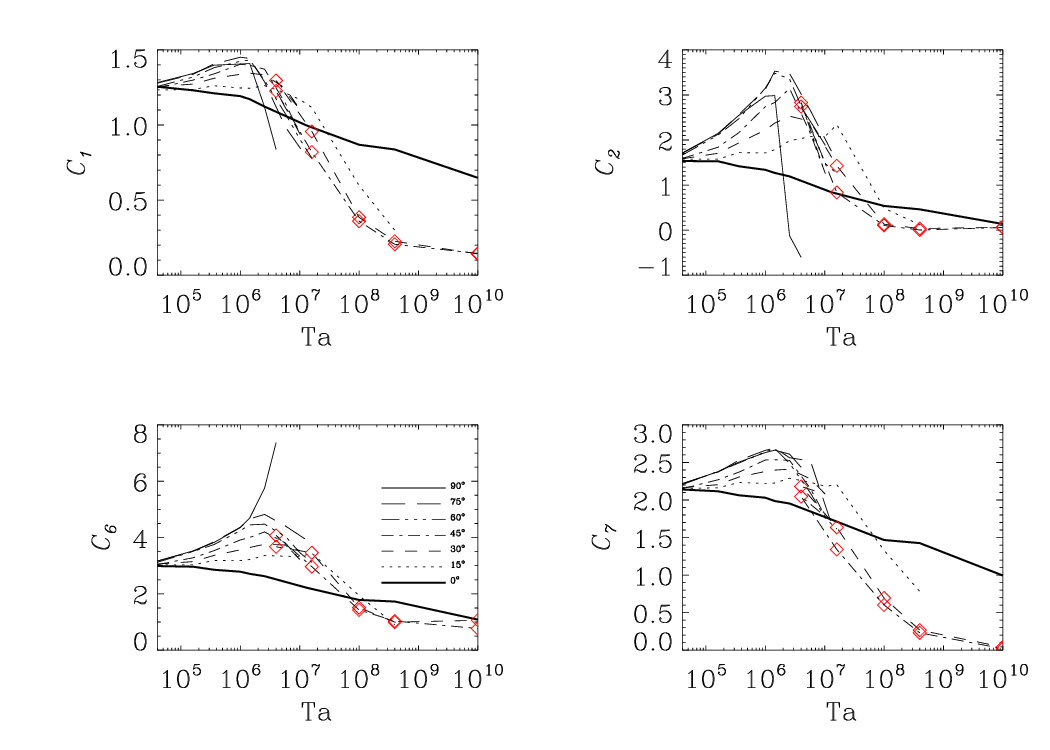}
\includegraphics[width=.75\textwidth]{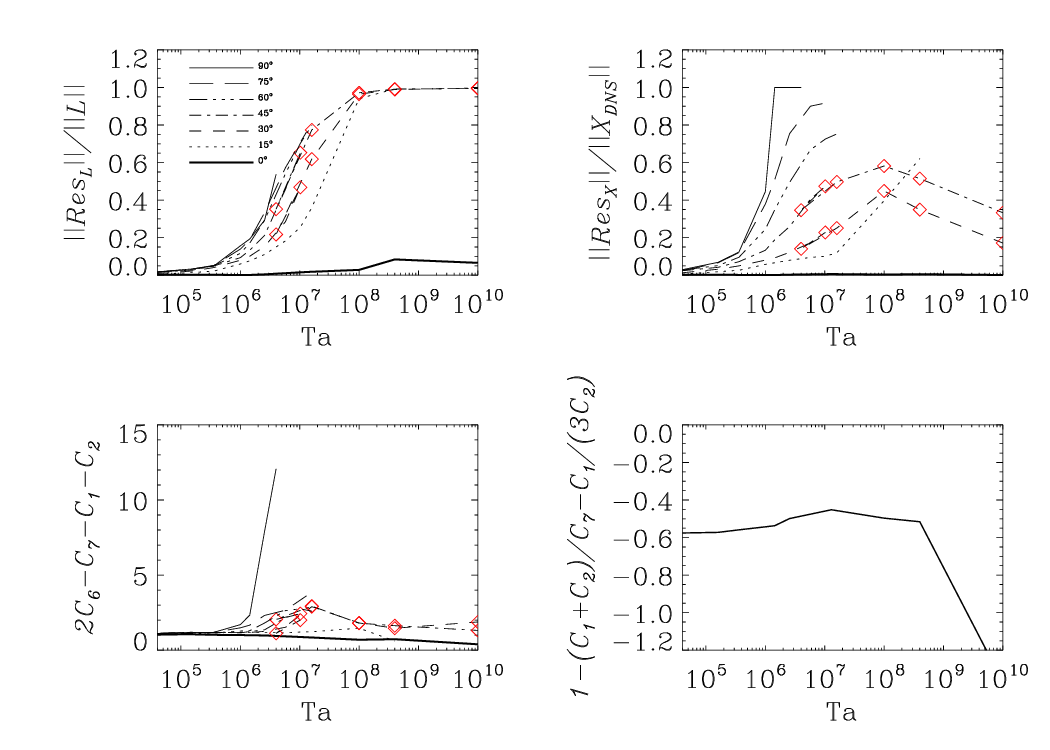}
\caption{\label{fig:lsvt1} Closure parameters from the least-squares
  approach. Upper four panels: $\{C_i\}$ as functions of $\Tay$;
  Third row of panels: normalized
  residuals \eqref{eq:eqres}
  and \eqref{eq:solres}. Lowest two panels: values of the constraint
  \eqref{eq:stcon} and 
  left-hand side of stability constraint \eqref{eq:stconB}
  (for $\vartheta=0$ only).
Symbols: data points for viscous heating included.
  \label{lspar}}
\end{figure*}

\begin{figure*}
\centering
\includegraphics[width=\textwidth]{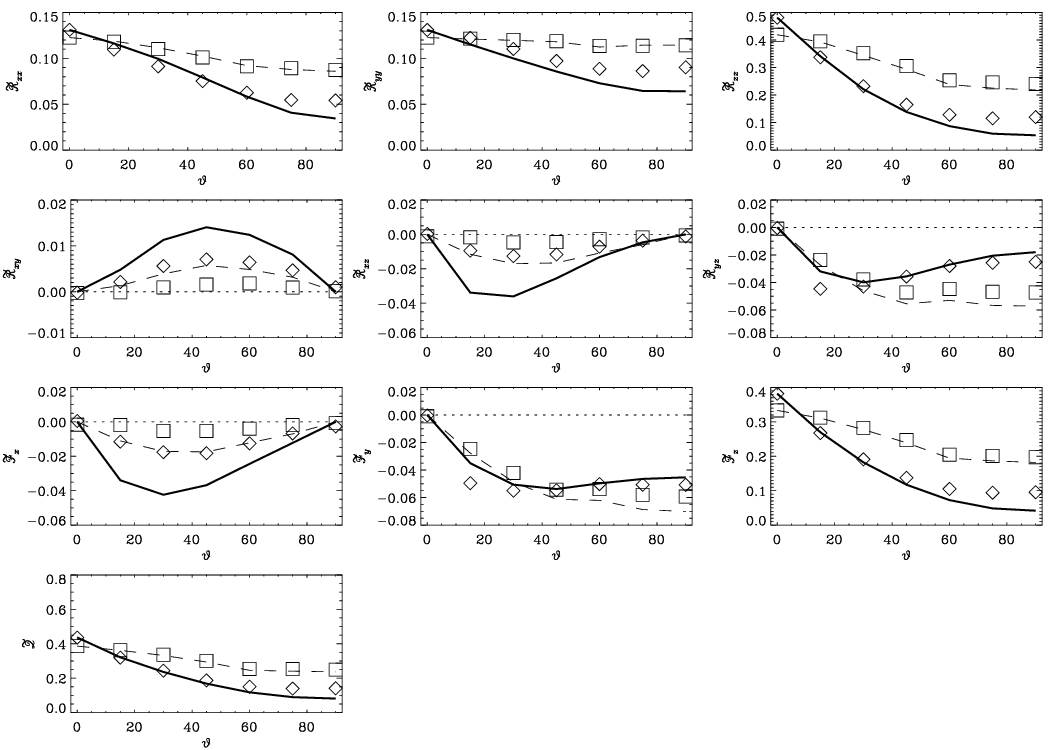}  
\caption{\label{fig:lsvt} 
Results of the closure  model
(lines) for the quantities $\meanRR$, $\meanFF$ and $\meanQ$
with
coefficients $\{C_i\}$ from the least squares fit compared with the 
corresponding DNS results
(symbols).
Reynolds stress, heat flux and temperature variance are normalized  by $U_0^2$, $d\, U_0 G_0$ and $(d\,G_0)^2$, respectively, with $U_0=d (\alpha g  G_0)^{1/2}$.
Note that the $\{C_i\}$ depend on rotation rate  $\Omega_0$
and colatitude $\vartheta$, see Fig.~ \ref{lspar}. Dotted lines/squares: 
 slow rotation (Runs A2--G2 with $\Tay=1.6 \cdot 10^5$),
solid lines/diamonds:
faster rotation (Runs A4--G4 with  $\Tay= 10^6$). 
The error bars associated with the DNS runs are not 
plotted since their size is, 
at most
of the same order of magnitude as that of the symbols.
}
\end{figure*}
  The DNS results are 
  qualitatively
  compatible with the specific
  solution \eqref{eq:specpole1}--\eqref{eq:specpole2}
  at the pole for all rotation rates considered
  insofar to good accuracy $\meanR_{xy,xz,yz}, \meanF_{x,y}=0$, 
  see Table~\ref{tab:bres} in the Online Material.

\subsection{Calibration of the closure model}

In this section we assume 
at first
that 
the diffusive coefficients
$C_{\nu,\nu\chi,\chi}$ vanish.
However,
they
can be thought to be subsumed by 
the coefficients $C_{1,6,7}$, see Eq.~\eqref{G1eq},
and 
we will
make an effort to
disentangle 
them
in Sec.~\ref{sect:dray} in the context of 
Rayleigh number dependence.

Stationary solutions of the closure model Eqs.~\eqref{eq:zdclo} result
for given $\{C_i\}$ from the corresponding nonlinear {\em algebraic}
system of equations. This opens up systematic ways of calibrating the
parameters and of studying thereby the performance of the ``minimally
extended GOMS10'' model.  Two such methods are described here.

\subsubsection{Least squares fit}

One can ask, whether any set of closure parameters 
$C_{1,2,6,7}$
could be found so that the 
stationary results from the closure model
reproduce
exactly
 the results from a
statistically stationary stage of a
 corresponding
 DNS run. 
A straightforward way to check this is to insert 
the parameters ($\alpha g$, $\Omega_0$, $G_0$)  used in the DNS together with their 
(temporally averaged)
results for 
$\meanRR$, $\meanFF$ and $\meanQ$ into the time-independent version of the system \eqref{eq:zdclo}. Treating 
the
$\{C_i\}$
as unknown variables, 
a generally overdetermined system of linear 
equations for them 
(ten equations vs. four variables) of the form 
\begin{equation}
\ve{N}\ve{c} = \ve{L}
\label{eq:lsprin}
\end{equation}
is obtained
where $\ve{c}=(C_1,C_2,C_6,C_7)$. 
The matrix $\ve{N}$ is derived from the 
closure terms and the vector $\ve{L}$ contains all remaining terms,
such as the
Coriolis and buoyancy terms, see Eqs.~\eqref{eq:zdclo}. 
Because of 
the overdetermination 
one can
in general not  expect to find any set of coefficients $\{C_i\}$ with which the closure 
reproduces all the modeled quantities perfectly.
At the pole, however, 
there are only four linearly independent equations in the system
\eqref{eq:lsprin}, 
making it 
unambiguously solvable. 
The solution is given explicitly
in Appendix~\ref{app:eq0}.
At all other latitudes 
this system can be solved only
approximately
using the standard linear least squares 
method, that is, solving the regular system 
$\ve{N}^{\rm T}\ve{N}\ve{c} = \ve{N}^{\rm T}\ve{L}$, where the superscript 
``T" denotes transposition. The results 
for $\ve{c}$
 are shown 
in Fig.~ \ref{lspar} as functions of the Taylor number 
and colatitude. 
As a test of the consistency of the Eq.~ \eqref{eq:lsprin}  
we 
calculated the residual norm
\EQ
{\rm Res}_{\ve{L}} = ||\ve{N}\ve{c}_{\rm ls}-\ve{L}||,  \label{eq:eqres}
\EE
where the subscript ``ls" refers to the least-squares solution and $||\!\cdot\!||$ denotes the Euclidian norm,
see  Fig.~\ref{lspar}. 
The
quality of the obtained solution can 
also be measured
by
calculating the 
difference in 
 the stationary solutions for
$\ve{X}=(\meanR_{xx}, \ldots, \meanR_{zz}, \meanF_x,  \meanF_y, \meanF_z, \meanQ )$ 
from the closure model \eqref{eq:zdclo} with $\ve{c}=\ve{c}_{\rm ls}$ 
and from the corresponding temporally averaged
results of the DNS run, that is, by calculating the residual
 \EQ
{\rm Res}_{\ve{X}} = ||\ve{X}_{\rm closure}-\ve{X}_{\rm\!DNS}||. \label{eq:solres}
 \EE
This quantity is also shown in Fig.~\ref{lspar} together with the left hand 
sides of the realizability condition 
\begin{equation}
2C_6-C_7-C_1-C_2 \ge 0
\label{eq:stcon}
\end{equation}
given in GOMS10 and the 
stability condition
\begin{equation}
1- \frac{C_1+C_2}{C_7} - \frac{C_2}{3C_1} 
<
0
\label{eq:stconB}
\end{equation}
for the pole as derived in the Appendix~\ref{app:eq0}.
The solutions $\ve{X}_{\rm closure}$ and $\ve{X}_{\rm\!DNS}$ are
directly compared in Fig.~\ref{fig:lsvt}.

From Fig.~\ref{lspar} one can see that the derived model coefficients 
change with colatitude and Taylor number, with following patterns: at the pole 
they fall with growing 
$\Tay$,
but not so for any other colatitude. Instead,  
they first grow with $\Tay$ and plateau or fall
for $\Tay>10^6$.
Both the growth and the fall become steeper with growing colatitude, 
and all the curves converge as $\Tay$ approaches zero.
Although the least squares method has no built-in way of adhering to 
the conditions \eqref{eq:stcon} and \eqref{eq:stconB}, we see that these 
are fulfilled nevertheless.

Using
the (exact) results for the non-rotating 
run Z 
we computed 
the different ratios of the coefficients $\{C_i\}$ and compared 
them to the corresponding ratios from GOMS10, S12a, and S12b
in Table~\ref{tab:cir}. 
Also listed in the Table are the ratios resulting from the 
non-rotating higher Rayleigh number runs that will be discussed in 
Sec.~\ref{sect:dray}.
 These ratios are
important because any
difference in the non-rotating case in the results for $\{C_i\}$
could be due to a badly chosen lengthscale $L$, which is canceled 
by the ratios. In any case, we see that our values are at odds with 
those of
GOMS10.
As for the residuals, they unsurprisingly vanish (as long as $\Tay \approx 2 \cdot 10^6$) at the pole, 
otherwise rise with Taylor number while the colatitude has only a small effect 
on the residual \eqref{eq:eqres}, but a stronger one on \eqref{eq:solres}.
The reason why the residuals do not vanish as expected at the pole for high Taylor numbers turns out to be the small deviations of the 
DNS data from what is theoretically expected from the system. For example, 
the off-diagonal Reynolds stresses $\meanR_{xy,xz,yz}$ and the temperature 
fluxes $\meanF_{x,y}$ are not exactly zero as expected. This is due to 
the fact that very long time integrations are needed for the time-averaged 
quantities to converge for highly fluctuating quantities.
The effect of this discrepancy, however, is minor when compared to the effect of having $\meanR_{xx}$ and $\meanR_{yy}$ diverge. On grounds of symmetry, $\meanR_{xx}$ and $\meanR_{yy}$ should be equal at the pole, 
but as can be seen in Table~\ref{tab:bres}, the DNS results for them are 
somewhat different in the pole at high rotation rates. This is due to the same
reason as for the off-diagonal stresses and horizontal heat fluxes. The 
least squares method can only produce perfect match with the DNS if the 
amount of linearly independent closure equations equals the number of model parameters, and both of the aforementioned deviations disturb this equivalence, something to which the least squares method seems to be 
sensitive.
At the equator the residual ${\rm Res}_{\ve{X}}$ 
settles at unity
for $\Tay \gtrsim 5 \cdot 10^7$. 
This is because the analytical results from the closure become very small for all 
the modeled quantities
indicating that the obtained $\{C_i\}$ do not allow for any other than the
trivial solution
of \eqref{eq:zdclo}.

\begin{table}
  \centering
  \caption[]{ \label{tab:cir} Ratios of the coefficients $\{C_i\}$ obtained from the non-rotating  
runs with different Rayleigh numbers  
compared to GOMS10 (G) and S12a,b.
}
  \vspace{-0.5cm}
  $$
  \begin{array}{p{0.06\linewidth}cc@{\hspace{1mm}}ccccccrrr}
    \hline
    \hline\\[-2.5mm]
    \noalign{\smallskip}
      \rm Run & \Ray/10^6 & C_1/C_2 & C_1/C_6 & C_1/C_7 & C_2/C_6 & C_2/C_7 & C_6/C_7 \\ 

    \hline\\[-2mm]
       \rm  Z & 0.30   & 0.81       & 0.42 & 0.58 & 0.51 & 0.72 & 1.40 \\
     \noalign{\smallskip}
    \hline\\[-2mm]
%       \rm R1 & 10  & 0.92233965 &      0.45319342 &      0.67777719 &      0.49135199 &      0.73484554 &       1.4955583 \\
       \rm R1 & 0.25        & 0.92       &      0.45 &      0.68 &      0.49 &      0.73 &       1.50 \\
    \noalign{\smallskip}
%       \rm R2 & 40  & 0.73308431 &      0.38836987 &      0.63976541 &      0.52977518 &      0.87270373 &       1.6473096 \\
       \rm R2 & \phm1   & 0.73       &      0.39 &      0.64 &      0.53 &      0.87 &       1.65 \\
    \noalign{\smallskip}
%    \rm R3 & 250 & 0.68209817 &      0.35037538 &      0.59077995 &      0.51367295 &      0.86612160 &       1.6861343 \\
       \rm R3 & \phm4   & 0.68       &      0.35 &      0.59 &      0.51 &      0.87 &       1.69 \\
    \noalign{\smallskip}
%       \rm R4 & 100  & 0.79525086 &      0.36350975 &      0.58058524 &      0.45710072 &      0.73006552 &       1.5971655 \\
       \rm R4 & \phm10      & 0.80       &      0.36 &      0.58 &      0.46 &      0.73 &       1.60 \\
    \noalign{\smallskip}
%       \rm R5 & 250  & 0.83008349 &      0.36368009 &      0.56684344 &      0.43812471 &      0.68287521 &       1.5586320 \\
       \rm R5 & \phm25      & 0.83       &      0.36 &      0.57 &      0.44 &      0.68 &       1.56 \\
    \noalign{\smallskip}
%       \rm R6 & 200 & 0.90172624 &      0.37004399 &      0.54515636 &      0.41037288 &      0.60456969 &       1.4732204 \\
       \rm R6 & 200         & 0.90       &      0.37 &      0.55 &      0.41 &      0.60 &       1.47 \\
    \noalign{\smallskip}
   {\rm S12a} & -           & \approx0.4 & -         & -         & -         & -         & -    \\
    \noalign{\smallskip}
%    $C_2/C_7$  & 0.72  & 0.87 & 0.87 & 0.68 & - & - & 0.43 \\
   {\rm S12b} & -           & -          & -         & -         & -         & -         & \approx2.1 \\
    \noalign{\smallskip}
%    $C_6/C_7$  & 1.40  & 1.65 & 1.69 & 1.56 & - & \approx2.1 & 1.00 \\
% {\rm GOMS10} & \phm50      & 0.66       & 0.29      & 0.29      & 0.43      & 0.43      & 1.00 \\ 
 {\rm G} & \phm50      & 0.66       & 0.29      & 0.29      & 0.43      & 0.43      & 1.00 \\ 
    \noalign{\smallskip}
    \hline
   \end{array}
   $$ 
\end{table}

The DNS results $\ve{X}_{\rm DNS}$ for a slow and a rapid rotation case
are compared with the corresponding closure results obtained with the 
derived model coefficients in Fig.~\ref{fig:lsvt}. 
At 
slow rotation
the DNS and the closure 
results are visibly closer to each other than at rapid rotation. 
Further,
the fit for small quantities
like $\meanR_{xy,xz,yz}$ or $\meanF_{x,y}$
 is 
significantly worse than for the large ones,
like $\meanR_{zz}$, $\meanF_{z}$ or $\meanQ$.

\subsubsection{Optimization approach}
\label{sec:opt}
\begin{figure*}
\centering
\includegraphics[width=.85\textwidth]{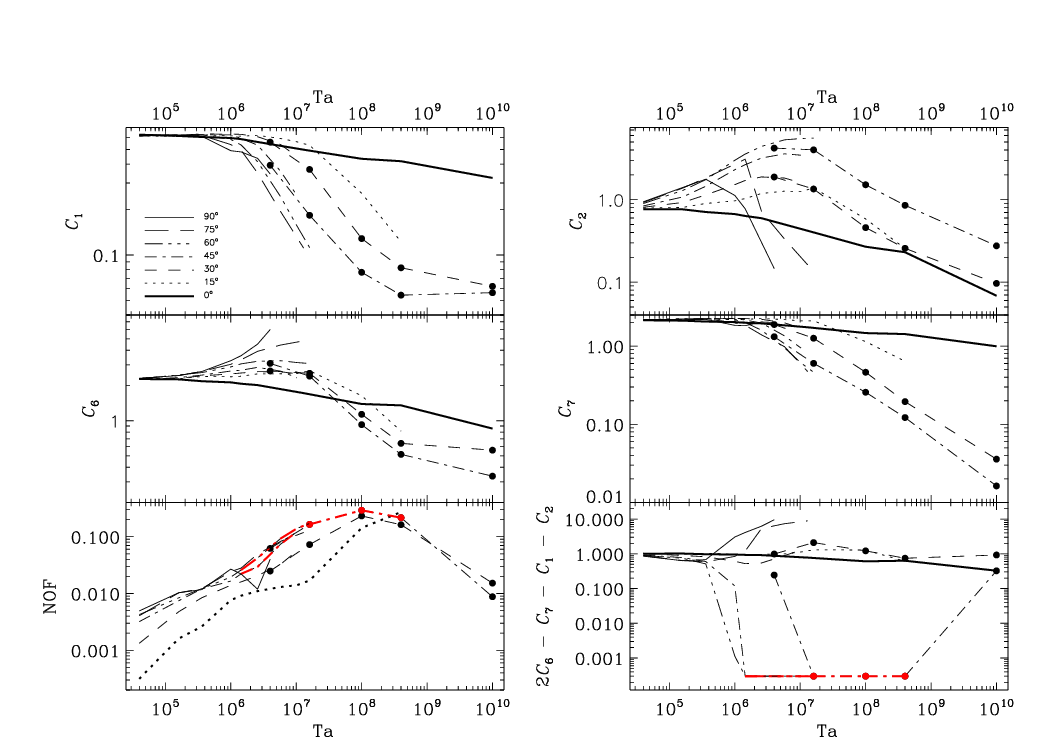}
\caption{\label{fig:Calt_stab} Closure parameters from the
  optimization approach for $\Ray=3\cdot 10^5$ and $\Pra=0.6$.
Upper four panels: $\{C_i\}$ as functions of $\Tay$
 and colatitude $\vartheta$.
Lower left panel: normalized objective function \eqref{eq:objnorm} at optimum
(NOF); values for $\vartheta=0^\circ$ are omitted because they reflect only roundoff errors.
Lower right:
value of 
$2C_6 - C_7 - C_1 -C_2$, 
see
\eqref{eq:stcon}.
 For legibility (numerical) zero values were replaced by an arbitrary small constant.
Red
curve sections: Regions where this constraint is active.
Symbols: data points for viscous heating included.
}
\end{figure*}

The least squares approach clearly shows that the stationary closure 
equations become increasingly inconsistent with the DNS 
away from 
the pole as the 
rotation rate is increased. However, as a method of calculating the 
$\{C_i\}$, it is inflexible: 
the matrix 
$\ve{N}$ and vector 
$\ve{L}$ are determined by the DNS-results inserted into them, and the method effectively minimizes ${\rm Res}_{\ve{L}}$ rather than ${\rm Res}_{\ve{X}}$. 
In order to determine the model coefficients while improving the agreement between the closure and the DNS results, we formulate the following 
optimization problem:
minimize the {\em objective function}
\EQ
        (\ve{X}_{\rm closure}-\ve{X}_{\rm\!DNS})^2 = {\rm Res}_{\ve{X}}^2 \label{eq:obj}
\EE
obeying
the constraints
$C_{1,2,6,7} > 0$ and \eqref{eq:stcon}. 
The problem was 
tackled
by the Generalized Reduced Gradient Method
\citep{LWJR78} as implemented by the IDL routine CONSTRAINED\_MIN
with the nonlinear system from Eqs.~\eqref{eq:zdclo} 
being solved by
Newton iteration (IDL routine NEWTON). The optimum results are shown
in Fig.~\ref{fig:Calt_stab} where the upper four panels refer to the
$\{C_i\}$ while the lower left panel gives the normalized objective
function 
\EQ 
(\ve{X}_{\rm closure}-\ve{X}_{\rm\! DNS})^2/\ve{X}_{\rm\!DNS}^2 \label{eq:objnorm} 
\EE 
and the lower right one shows the
value of the quantity $2C_6 - C_7 - C_1 -C_2$ from the constraint
\eqref{eq:stcon}.  Along with the dependence on $\Tay$, there is again
in general a separate one on $\vartheta$.  At the pole ($\vartheta=0$)
the objective function assumes exceptionally low values (not shown).
This is a consequence of the already mentioned degeneration which
allows to determine the $\{C_i\}$ uniquely from the $\ve{X}_{\rm
  DNS}$.  Hence the objective function actually vanishes and the
observed values are completely due to the iterative nature of the
solution and roundoff errors.  Apart from the pole, the quality of the
optimum is in general decreasing with growing $\Tay$, yet having only
a weak dependence on colatitude.  It can be considered good up to
  $\Tay=\mbox{a few times}\: 10^6$ or $\vartheta\leq15^\circ$ and 
again for $\Tay
  \gtrsim 10^9$. For $10^6 \le \Tay\le1.6\cdot10^7$ and $\Tay$
dependent $\vartheta$ intervals between $45^\circ$ and $75^\circ$, the
constraint \eqref{eq:stcon} becomes ``active" in the sense that the
optimum lies then on the margin of the admissible domain, that is, $2
C_6 - C_1 -C_2 - C_7=0$ (red curve sections in
Fig.~\ref{fig:Calt_stab}). For $\Tay\approx 10^8 \ldots 10^9$,
  $C_1$ and $C_7$ show in general a monotonously falling dependence on
  $\Tay$ which is weak as long as $\Tay\lesssim 10^6$. The dependences
  resemble Lorentzians at least for $\Tay \lesssim 10^7$ and
  $\vartheta\ge 45^\circ$. In these ranges we find also a weak dependence
  on $\vartheta$.  Beyond $\Tay\approx 10^8 \ldots 10^9$ the two
  coefficients start to grow again at 
  lower
  latitudes.  In contrast,    
  $C_2$ and $C_6$ are falling monotonously with $\Tay$ only at the
  pole. At larger $\vartheta$ the $\Tay$ dependences show maxima in the
  interval $10^6 < \Tay < 10^7$ (with some exceptions with saturating
  behavior). In all, the behavior of the latter two coefficients seems
  less systematic than that of $C_1$ and $C_7$.

By comparing the normalized objective function (NOF) in Fig.~\ref{fig:Calt_stab}) and the 
(square of the)
residual ${\rm Res}_{\ve{X}}$ in Fig.~\ref{lspar} one sees 
that although they depict the same 
deviation, NOF does not become as great as ${\rm Res}_{\ve{X}}$ 
for
larger Taylor numbers.
This means that the optimization procedure succeeds in finding closure parameters with better matching results for $X$, as intended.

Given that, apart from the constraint \eqref{eq:stcon}, also the stability properties of the closure model should not differ from that of the DNS,
we 
enhanced the optimization problem 
by the constraint that for the optimum fit the stationary 
solution corresponding
to it should be stable. For that, we linearized the system \eqref{eq:zdclo} about the state $\ve{X}_{\rm closure}$, obtaining a system of the form $\partial_t (\ve{\delta X}) = \ve{A} \cdot \ve{\delta X}$ for the 
perturbations $\ve{\delta X}$, and required that the
maximum of the real parts of the eigenvalues of $\ve{A}$ is
negative. To avoid influences of numerical noise we set their
upper bound to a small negative value instead of zero.
The matrix eigenvalue problem was solved by means of the IDL routines LA\_ELMHES and LA\_HQR.
It turned out that the additional constraint is never active,
that is, that stability is already granted if \eqref{eq:stcon} is
obeyed.

\begin{figure*}
\centering
\includegraphics[width=\textwidth]{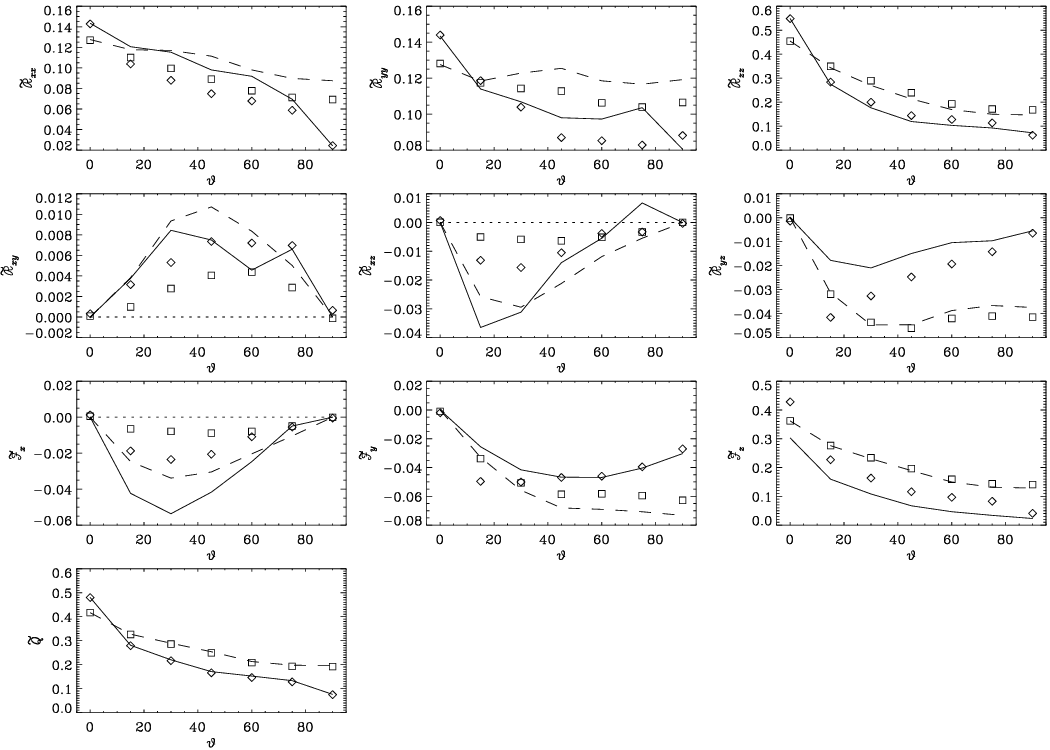}
\caption{\label{fig:X1}
Same as Fig.~\ref{fig:lsvt} but with $\{C_i\}$ from the optimization approach
%MR: added
and for $\Tay=3.6\cdot 10^5, 2.56 \cdot 10^6$.
}
\end{figure*}

In Fig.~\ref{fig:X1} the DNS and closure model results as functions of
the colatitude are given for the same two Taylor numbers as
  in Fig.~\ref{fig:lsvt}.  Obviously, the dominant
variables $\rzz$ and $\meanQ$ are very well fitted whereas
  $\meanF_z$ is less accurate for the higher rotation rate. For slow
rotation, also the other quantities except $\rxy$, $\rxz$ and
$\meanF_x$ show good fits.  At the higher rotation rate, the
  quantities of intermediate magnitude, $\meanR_{xx,yy,xz,yz},
  \meanF_{x,y}$ show different fit qualities.  In all, we have to
  conclude that the incompleteness of the closure ansatz is most
  clearly visible
  in the quantities $\meanR_{xz,yz},
  \meanF_{x,z}$ while there is apparently nothing important missing in
  the ansatzes for $\meanR_{zz}$ and $\meanQ$. Thus, a guideline is
  found how to improve the ansatzes with the added terms having
  maximum effect. 
 
Both versions of the optimization approach produce up to roundoff
errors identical results as long as neither of the two constraints is
active.
The dependence of the objective function on latitude is in general
weak, and beyond $\Tay\approx 10^7$ its dependence on $\Tay$ is weak
too.
Again, the ratios $C_1/C_2$ and $C_6/C_7$ for the non-rotating case are at odds with GOMS10, see Table~\ref{tab:cir_opt}.
\begin{table}
  \centering
  \caption[]{ \label{tab:cir_opt} Ratios of the coefficients $\{C_i\}$ obtained 
  for rotating runs by the optimization approach. $\Tay= 4\cdot10^4 \ldots 10^8$}  
  \vspace{-0.5cm}
  $$
  \arraycolsep10pt
  \begin{array}{rccc}
    \hline
    \hline\\[-2.5mm]
\vartheta \:[^\circ]    &C_1/C_2    &  C_1/C_6       &    C_1/C_7 \\[1mm]
    \hline\\[-2.5mm]
   0          & 0.81 - 1.73   &  0.42-0.52          &  0.58 - 1.45 \\
  15        &  0.55 - 0.88 &  0.29-0.42         &  0.55 - 0.88 \\
  30        &  0.52 - 0.81 &  0.21 -0.42        &  0.52 - 0.81 \\
  45        &  0.54  - 0.83 &  0.13-0.42       &  0.51 - 0.83 \\ 
  60        &  0.51 - 0.89 &  0.06-0.42       &  0.51 - 0.89  \\
  75        &  0.49   - 0.89   &  0.01-0.42    &  0.49 - 0.89 \\
  90        &  0.43  - 1.11&  0.01-0.42    &  0.43 - 1.11\\[1mm]
    \hline\\[-2.5mm]
    \end{array}
    $$
    \end{table}
When comparing the results of the least-squares and the optimization approaches we find major quantitative differences in $ C_{1,2,7}$, 
in addition differing monotony in $C_{1,7}$ while the residuals are
clearly smaller for the latter approach, namely $\approx30\%$ vs. up
to 100 \% for the former.

\subsubsection{Dependence on $\Tay$}
For both approaches, 
the obtained $\{C_i\}$ show a 
clear
 dependence on $\Tay$. 
As long as the fit quality is satisfactory, that is for 
$\Tay$
up to a few times $10^6$
for which the effect of the neglect of closure terms constructed from $\OO$ should be small (see Sec.~\ref{sec:closure}),
and again for $\Tay \gtrsim 10^9$
these dependencies might be taken as physical, but we are faced 
with the ambiguity between the two fitting approaches.
At the pole, however, the fit quality is perfect and no ambiguity occurs. As discussed above,
the used closure ansatzes are here complete at least up to the level represented by \eqref{eq:extR}, \eqref{eq:extF}.
Hence, the 
$\Tay$
dependences of the $\{C_i\}$ are here the more trustworthy. 
On the one hand, the DNS results are  consistent with the 
specific solution \eqref{eq:specpole1}--\eqref{eq:specpole2} for the
pole, which was derived from the
closure model, but on the other hand they clearly depend on $\Omega$ while there is no explicit 
occurrence of $\OO$ in the specific solution. 
We interpret this as a confirmation
of our statement in Sec.~\ref{sec:closure} that the closure should be
extended by making the coefficients $\Omega$ (or $\Tay$) dependent.
Apart from the pole, the quality of the fit 
from the optimization approach
is gradually worsening with increasing rotation rate
up to $\Tay\approx 10^8$, but improves again beyond that letting the NOF adopt values $< 10^{-3}$ for $\Tay\approx 10^{10}$.
We take this as an indication of
a most pronounced 
importance of closure terms constructed from $\OO$
being structurally different from the original ones
for medium rotation rates $10^6 \lesssim \Tay \lesssim 10^9$.
In this range
 the $\Omega$ (and $\vartheta$) dependence
induced by 
those terms
cannot be adequately ``mimicked" by corresponding dependences of the $\{C_i\}$.
In contrast, for slow and rapid rotation the original ansatz performs
satisfactorily well.
 
\subsubsection{Dependence on box aspect ratio}
\label{sec:aspectr}

The elevator modes are only weakly damped for a box of aspect ratio
unity in the non-rotating case (or at the pole).
Hence, in order to assess their influence (and that of possibly existing similar weakly
damped modes at other latitudes)
we performed a series of runs for the chosen set of
latitudes with $\Gamma = L_z/ L_x = L_z/L_y = 4$, $\Ray=3\cdot10^5$
and a moderate rotation rate of $\Tay\approx4\cdot10^9$ or $1.6\cdot
10^7$ defined with $d$ in \eqref{eq:Tay} taken as the $z$ or $x=y$
extent of the computational box, respectively.
The runs are listed in Table \ref{tab:bresGam}.

\begin{figure*}
\centering
\includegraphics[width=1\textwidth]{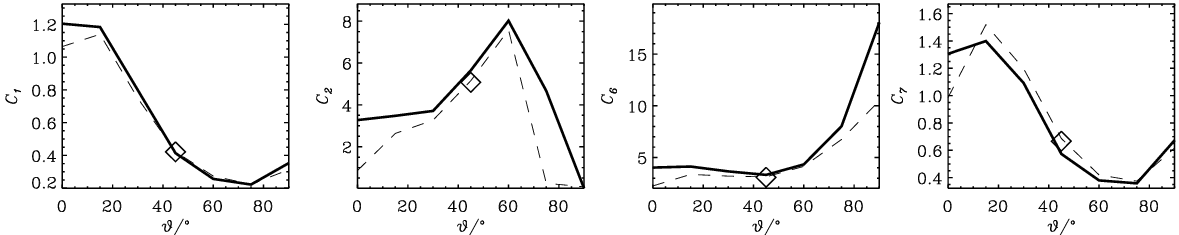}
\caption{\label{fig:gamdep} Closure coefficients $\{C_i\}$ for box aspect ratios $\Gamma=4$ (solid) and $\Gamma=1$ (dashed) both with $\Ray=3\cdot10^5$, $\Tay=1.6\cdot 10^7$
  (defined with the horizontal box extent) as functions of colatitude
  $\vartheta$. Symbols: values for $\Gamma=0.75$ for the same $\Tay$.}
\end{figure*}

The results for the $\{C_i\}$, obtained by the optimization-based fit,
are shown in Fig.~\ref{fig:gamdep} in combination with results for
$\Gamma=1$ where comparability was ensured by equating the Taylor
numbers defined with the horizontal rather than the vertical extent of
the box. Apart from the pole and with an exception for $C_2$ at
$\vartheta=75^\circ$, there is obviously only a weak influence of the
aspect ratio which supposedly does not exceed the general
uncertainties in the determination of the $\{C_i\}$.  We interpret the
systematically stronger deviations at the pole as an indication of
changes in the overall statistical properties of the turbulence due to
the $\Gamma$ dependent damping of the elevator modes.  Such changes
are visible in the time series of the quantities $\vec{X}$: For
$\Gamma=1$ both the temporal averages and the magnitudes of the
temporal fluctuations are clearly higher than for $\Gamma=4$. The
frequency of sharp high-amplitude peaks is higher in the former case
and the time series has a somewhat clearer quasiperiodic character.
The large deviation in $C_6$ at the equator cannot be
  explained by the occurrence of elevator modes given that the
  Rayleigh number is clearly subcritical
for them.

As a spot check we also performed one run with $\Gamma=0.75$ (one of
the cases considered in GOMS10) and $\vartheta=45^\circ$,
$\Tay\approx5\cdot10^6$. Again a stationary state was reached albeit
with an even stronger quasi-periodicity in its time series compared to
$\Gamma=1$.  The coefficients are very close to those obtained with
$\Gamma=1$ and hence also to those for $\Gamma=4$, see the symbols in
Fig.~\ref{fig:gamdep}.  We conclude that at not too small rotation
rates and colatitudes the influence of the box aspect ratio on the
coefficient values is not important.

\subsubsection{Dependence on $\Ray$}
\label{sect:dray}
\begin{figure}
\centering
\hspace*{-10mm}\includegraphics[width=.59\columnwidth]{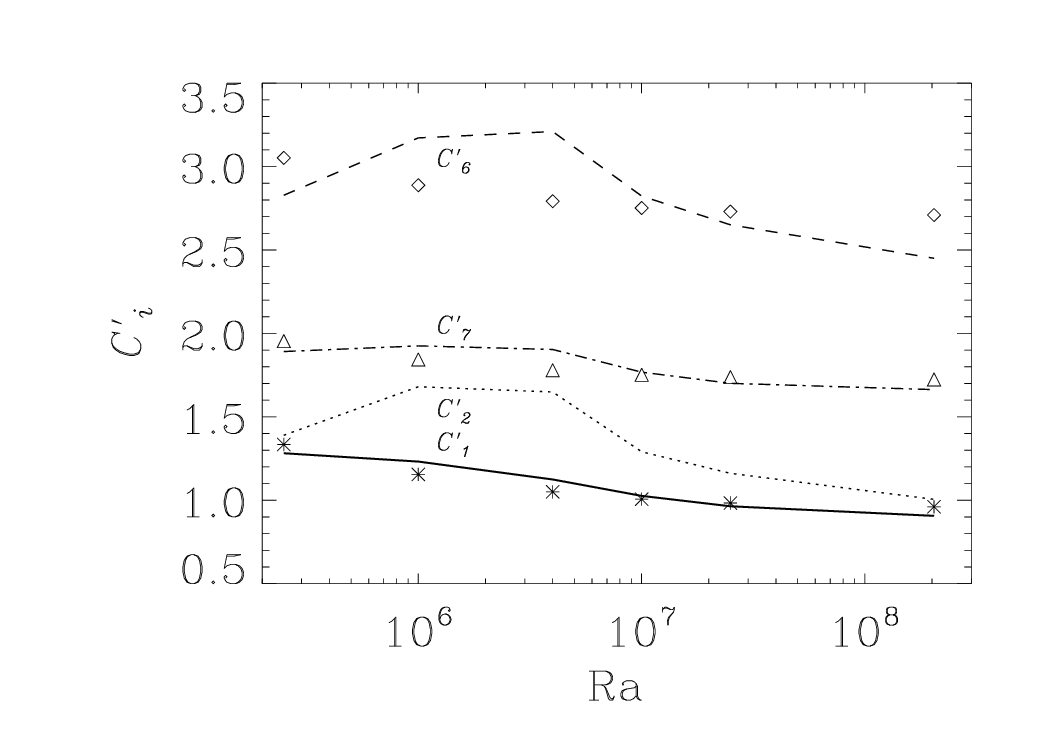}\hspace{-8mm}
\includegraphics[width=.59\columnwidth]{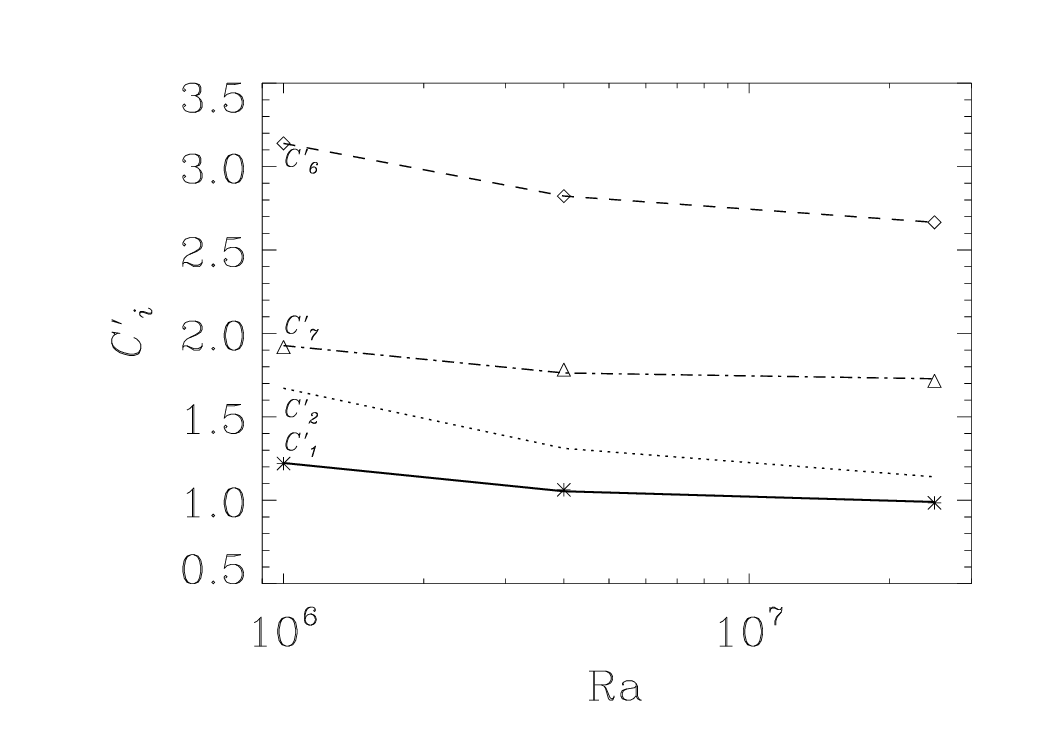}\hspace{-8mm}
\caption{\label{fig:G2x}
Closure parameters from the least-squares fit (lines)
in the cases $\Tay=0$ (left) and $\Tay=10^6$ with $\vartheta=0$ 
(right) as functions of $\Ray$
with $\Pr=1$.
Corresponding closure parameters $C_{1,6,7}'$ (symbols) 
obtained from \eqref{G1eq} with the values \eqref{eq:unival}.
}
\end{figure}

Due to computational constraints that arise as a consequence of
the required higher resolution, we have not studied Rayleigh numbers
higher than 
$\Ray = 3 \cdot 10^5$
in detail.
This is relevant because the comparison study, GOMS10, employed values up to two orders of magnitude higher.
In order to see how our results are influenced by the Rayleigh number 
we performed two sets of runs with higher $\Ray$ 
and
 $\Tay=0$ (up to $\Ray=2 \cdot10^8$) 
as well as
 $\Tay=10^6$, $\vartheta=0\degr$ (up to $\Ray=2.5 \cdot 10^7$).
Since both of these cases are solvable exactly, we use the least squares method to calculate the model coefficients.
The Rayleigh numbers were changed by adjusting the diffusivity parameters $\chi$ and $\nu$, while keeping 
the Prandtl and Taylor numbers constant.
The results are
summarised in Table~\ref{tab:simsetsB} and shown in
Fig.~\ref{fig:G2x}.
Table~\ref{tab:bres2} in the Online Material gives more details. The resulting ratios of the 
coefficients $\{C_i\}$ for the non-rotating case are also listed in Table~\ref{tab:cir}. 

From Fig.~\ref{fig:G2x} we see that the values of
the coefficients $C_2$ and $C_6$ rise somewhat with increasing Rayleigh number and then fall below their 
initial values, while the other coefficients fall monotonously. The ratios of the coefficients shown in 
Table~\ref{tab:cir} 
exhibit 
different behaviors with increasing $\Ray$, from monotonously falling 
($C_1/C_6$ and $C_1/C_7$) to first falling and then rising ($C_1/C_2$) and first
rising and then falling
(all others).

The parameters obtained from the rotating runs are rather similar 
to those from the non-rotating runs.
$C_{1,7}$ as functions of $\Ray$ 
tend to follow similar patterns
as obtained for 
the non-rotating 
runs, 
but $C_{2,6}$ are monotonously decreasing.
The 
results for the highest 
$\Ray=2.5 \cdot 10^7$ 
are
very close to each other for the non-rotating and rotating cases. This
is because the rotational influence on the flow, measured by $\Co$, is
  for constant $\Tay$
  decreasing
with increasing $\Ray$.
The behavior of
the $\{C_i\}$ from the rotating
cases suggests convergence to some constant values at high $\Ray$, but
even higher $\Ray$ runs would be
needed to verify this. In the non-rotating case 
a tendency towards convergence for higher values of $\Ray$
is apparently not yet reached.
In any case, the coefficients are not drastically changed by the
Rayleigh number.

Nevertheless it seems worth a try to remove the $\Ray$ dependence of the $C_{1,6,7}$ completely
by reinstating the 
diffusive
 terms parametrized by the $C_{\nu,\nu\chi,\chi}$ in the closure.
In doing so, we rename the $\Ray$ dependent $\{C_i\}$ obtained as described before under the assumption of vanishing 
diffusion
by labelling them with a prime and set according to Eqs. \eqref{eq:clor} -- \eqref{eq:cloq}
\begin{align}
C'_1 &= C_1 + \nu C_\nu/(L \sqrt{\meanR}), \nonumber  \\
C'_6 &= C_6  + \onehalf (\nu + \chi) C_{\nu\chi}/(L \sqrt{\meanR}), \label{G1eq}  \\
C'_7 &= C_7 + \chi C_{\chi}/(L \sqrt{\meanR}). \nonumber
\end{align}
We note that it is not possible to write a similar expression for $C_2$. Hence, it must here remain $\Ray$ dependent.

Given $N$ runs with different values of $\Ray$, the two closure
coefficients occurring in 
each
of the equalities \eqref{G1eq},
but now being assumed to be independent of $\Ray$, can be determined by a standard least-squares approach.
In the non-rotating case we obtain in this way
\EQ
\begin{alignedat}{6}
&C_1      &&= 0.9,      \quad &&C_6      &&=2.7,      \quad &&C_7&&=1.7,\\
&C_{\nu}&&= 164.7, \quad &&C_{\chi}&&=101.6, \quad &&C_{\nu\chi}&&=151.1 \:.
\end{alignedat} \label{eq:unival} 
\EE
For giving an impression of the quality of the fit, these values have been employed
in  \eqref{G1eq} to re-calculate the $C'_{1,6,7}$ and Fig.~\ref{fig:G2x} (left) shows the results in comparison with the original $\Ray$-dependent values.
In the $\Ray$ interval studied, the assumption of $\Ray$--independence seems well justified for $C_1$ and to a bit lesser degree also for $C_7$, but
not for $C_6$.
Figure \ref{fig:G2x} (right) presents corresponding data for the rotating case $\Tay=10^6$, $\vartheta=0$.
Here, the fit is much better, but we had only three data points to consider.

\subsubsection{Dependence of the Nusselt on the Rayleigh number}

The results for various Rayleigh numbers allow us to study the
dependence of the Nusselt number on $\Ray$ 
as it results
 from both the closure
model and DNS.  A similar exercise was done in GOMS10 for the
inhomogeneous case.

From the closed-form solution \eqref{eq:meanR} of the closure model one can derive the asymptotic behavior 
of $\rm Nu(\Ray)$ for $\Ray\rightarrow\infty$  in the cases $\vec{\Omega}=\vec{0}$ or $\vartheta=0$.
 Replacing the constants $C_{1,6,7}$ in \eqref{eq:meanR} according to \eqref{G1eq} and assuming the Prandtl
number to be finite and independent of $\Ray$, it can be seen that the only consistent assumption for $\meanR$ in this limit is $\meanR=\,$const. Then we have from \eqref{eq:Nu} and \eqref{eq:meanF}
the relation
$\Nu(\Ray) \sim \Ray^{1/2}$ for $\Ray\rightarrow\infty$ as expected.
The same scaling was obtained in \cite{CLTT05}.

In order to compare the closure model results with the numerical 
ones we need to require 
$C_{\nu,\nu\chi,\chi} \neq 0$. 
Otherwise, the diffusivities $\nu$ and $\chi$ would 
have no effect 
rendering the results same for all
$\Ray$. 
To calculate the Nusselt number from the closure we use the constant ($\Ray$--independent) values 
 for $C_{1,6,7,\nu,\nu\chi,\chi}$
derived in Section \ref{sect:dray}, 
Eq.~\eqref{eq:unival}
and take the average of the least squares results for $C_2$ illustrated in
Fig.~\ref{fig:G2x}.

\begin{figure}
\centering
\includegraphics[width=\columnwidth]{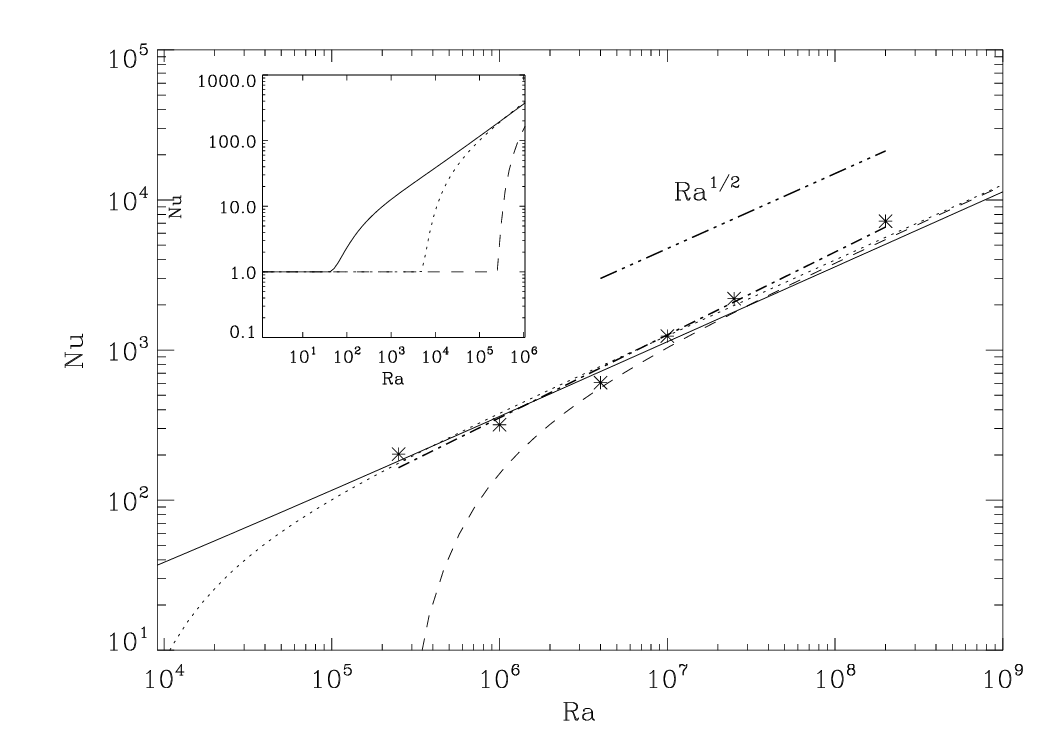}
\caption{\label{fig:Nura} 
Dependence of the Nusselt number on Rayleigh number,
$\Tay=0$, $\Pr=1$.
Symbols: DNS results. 
Results of the closure model with different model coefficients --  
solid line:  original 
GOMS10 values $C_{1,2,6,7}=0.4,0.6,1.4,1.4$, rescaled with
$\pi^{1/2}$, 
$C_{\nu,\chi,\nu\chi}=12,6,2$, rescaled with
$\pi$;
%JES: the C_{1,2,6,7} are multiplied by sqrt(\pi), C_{\nu,\chi,\nu\chi} by \pi, 
%because of the structure of the closure terms.
dotted: 
$C_{1,2,6,7}=0.9,1.4,2.7,1.7$,
$C_{\nu,\chi,\nu\chi}=164.7,101.6,151.1$ from \eqref{eq:unival},
with $C_2$ taken to be the average of the values in Fig.~\ref{fig:G2x};
dashed: 
same as dotted, but $C_{\nu,\chi,\nu\chi}=1000$ chosen arbitrarily.
Dash--dotted line:
power law
fit 
${\rm Nu} = 0.17 \Ray^{0.55}$ to the DNS data;
dash--triple--dotted: $\Nu\sim\Ray^{1/2}$ asymptotics.
}
\end{figure}

The resulting ${\rm Nu}(\Ray)$ dependence is shown in Fig.~\ref{fig:Nura} 
together with the relation obtained with the original GOMS10
closure coefficients.
To demonstrate the effect 
of
the diffusive terms 
 on the closure results, we have also plotted 
the results for the arbitrarily chosen values $C_{\nu,\chi,\nu\chi}=1000$. One can readily see 
that they contribute only in the low Rayleigh number regime,
especially altering the critical $\Ray$.
In the asymptotic regime,
the values obtained from DNS, plotted
with symbols, are in fair agreement with
our
closure results,
and with those of GOMS10.
(Note that there $L=L_x/\sqrt{\pi}$ was used in contrast to our choice $L=L_x$.
This effectively means a rescaling of the closure parameters.)

Also plotted are the asymptotic ${\rm Nu} \propto \Ray^{1/2}$
(dash-triple-dotted line) and an power law fit ${\rm Nu} \approx 0.17
\Ray^{0.55}$, obtained with linear regression.

\subsubsection{Reynolds stress and heat fluxes in comparison to
  compressible simulations}

The off-diagonal Reynolds stresses and the turbulent heat flux are
important in generating the differential rotation of stellar
convective envelopes \cite[e.g.][]{R89}. These quantities have been
computed from numerous simulations of compressible convection in
Cartesian \citep[e.g.][]{PTBNS93,C01,KKT04,REZ05} and spherical geometries
\citep[e.g.][]{RBMD94,KMGBC11}. It is important to compare the results of our
homogeneous Boussinesq runs to those in the literature in order to
draw conclusions on the robustness of certain features such as the
latitude and rotation rate dependence.

We find that $\rxy$, corresponding to latitudinal flux of angular
momentum is always positive, i.e.\ directed towards the equator in
accordance with previous DNS and analytical theory
\citep{KR93,KR05}. There is a tendency for the maximum of $\rxy$ to
move toward the equator as the rotation rate is increased in
accordance with compressible simulations of \cite{C01} and
\cite{KKT04}. Furthermore, the vertical flux corresponding to $\ryz$
is always negative. No sign reversal, observed at high $\Co$
in compressible runs of \cite{KKT04}, is seen even for the highest
Taylor numbers. The third off-diagonal component $\rxz$ is mostly
negative, although positive values occur at mid-latitudes for rapid
rotation. Earlier results suggest that positive values occur at high
latitudes only \citep[e.g.][]{PTBNS93}.

Apart from the equator
the latitudinal heat flux $\meanF_x$ is always directed towards the
pole;
 the azimuthal heat flux $\meanF_y$ is 
 always
 negative, i.e.\ in the
retrograde direction. These features are broadly in accordance with
Cartesian \citep[e.g.][]{REKK05} and spherical simulations
\citep[e.g.][]{KMGBC11}.
One puzzling feature of our simulations is the monotonously increasing 
$\meanF_z$ at the pole as a function of Taylor number, see Table    
\ref{tab:bres}. At 
colatitude
$15\degr$, $\meanF_z$ is monotonously
decreasing 
throughout, but
at all other colatitudes 
$\meanF_z$ 
has a minimum at 
$\Tay=1.44\cdot 10^6 \ldots 1.3 \cdot 10^7$.
We also note that $\rxz$, $\meanF_x$, and $\meanF_y$ obtain 
values of the order of $\rxx$ and $\meanF_z$,
for Taylor numbers $10^8\ldots 10^{10}$
and $10^{10}$, respectively,
at all latitudes except the pole and the equator. In some of these cases the flow structures
are rather laminar which may reflect the fact that convection is only
mildly supercritical.  %(So $\Tay=10^{10}$ is already subcritical, but at the pole.)
The large values of $\rzz$, $\meanF_z$ and $\meanQ$ at the pole for the highest $\Tay$
might be explainable by a strong dominance of the elevator modes if the secondary instabilities
which are limiting their growth are suppressed by rapid rotation.

\section{Conclusions}
\label{sec:conclusions}

The closure presented in GOMS10 has been known not to reproduce
essential
flow features under rotation, at least when the rotation and gravity vectors are 
aligned. In this study we
made an attempt to
extend the applicability of the model in the presence of rotation by allowing 
the model parameters to depend on the
rotation rate
(or  Taylor number). A similar modification to 
the GOMS10 model can be found in \cite{MG07}, 
where
the length scale $L$ was assumed to vary as a function of the rotation rate.
Our approach is more general because it allows
the model parameters
to obey individual dependences on  $\Tay$.

The main conclusion to be drawn from our investigations with the
homogeneous Boussinesq closure model is that
the extension described above 
works perfectly at the pole, while
elsewhere
the
validity of the closure degrades at first as rotation rate and colatitude
are increased, as indicated by the growing residuals of the parameter
fits (see Figs.~\ref{fig:lsvt1} and \ref{fig:Calt_stab}).
For even higher rotation rates, however, the closure validity recovers again.
This suggests
that 
even this modified
GOMS10 closure is essentially incomplete for intermediate rotation rates.
In particular, given the clear an\-iso\-tro\-py induced by the direction
  of rotation, the purely iso\-tro\-pi\-zing character of the closure ansatz
  should be revised.

However, even in the non-rotating case we were not able to reproduce
the ratios of the coefficients $\{C_i\}$ provided in GOMS10, see
Table~\ref{tab:cir}. 
Increasing the Rayleigh number in the DNS runs did not solve this issue.
This might be attributed to the fact that the coefficients in that study were not independently 
determined from DNS, but instead adopted without change from the inhomogeneous model while only 
adjusting the length scale $L$.

We observed that positivity of the parameters and their
adherence to the realizability condition alone
always
guarantees stability of the stationary solutions of the closure model
in accordance with the
stability of the underlying statistically stationary DNS
solution. 

In this study we have only briefly explored the 
influence
of the Rayleigh 
number on the optimum model coefficients. One motivation of this was to see, 
whether the parameters 
settle to some constant values with 
increasing
$\Ray$.
Apart from some weak signs of convergence in rotating 
runs,
we found no clear asymptotic tendencies, although the acquired parameters do not change dramatically. 
More systematic efforts
are
needed
to clarify this issue.
Future research 
should also extend
 the present work to 
 other
 settings including
bounded
domains
which require a
one-di\-men\-sio\-nal version of the closure
as already employed
in GOMS10.
Moreover,
from our inspections conclusions can be drawn
which amendments to the model would have the greatest benefit.

We also set out to investigate the effect of changing the aspect
  ratio of the computational domain. As a result, a weak dependence
  was found in any other latitudinal location than the pole. 
  This can be explained by the fact that
  the elevator
  modes, that are excited in the non-rotating case, are no longer seen
  away from the pole when rotation is applied. Even at the pole
  and with aspect ratio unity,
  these modes are subject to parasitic instabilities that eventually
  suppress them,
  having yet a noticeable influence on the turbulence.

The approaches used by \cite[e.g.][]{X1989,CM96,C1997,C2011} are 
sufficiently far from the GOMS10 closure so that a
comparison can hardly be performed with respect to their theoretical
bases, but rather a comparison of the results of the models with each
other and with DNS. However, such comparisons are not within the scope 
of this study.

\begin{acknowledgements}
  The computations were performed on the facilities hosted by the CSC
  -- IT Center for Science in Espoo, Finland, who are financed by the
  Finnish ministry of education, and on the FGI and Helsinki University 
  cluster 'Alcyone'. The authors acknowledge financial
  support from the Academy of Finland grant Nos.\ 136189, 140970
  (PJK), 218159 and 141017 (MJK), and the University of Helsinki
  research project `Active Suns'. The authors acknowledge the
  hospitality of NORDITA. JES acknowledges the financial support from
  the Finnish Cultural Foundation. 
We thank 
Elizabeth Cole for help in improving the language.

\end{acknowledgements}

\appendix

\section{Closure model equations for the inhomogeneous Boussinesq system}
\label{app:eq}

Here we describe the closure model for the 
general inhomogeneous case
of Boussinesq convection,
first without restricting to a specific mean.
Evolution equations for the Rey\-nolds stress and turbulent heat flux 
can be derived from the equations for the 
fluctuating
quantities $\uvec$ and 
$\theta$. They read
\begin{align}
\hspace*{-1mm}\frac{\partial \vec{u }}{\partial t} = &-\mean{\Uvec} \cdot \bm{\nabla} \vec{u }-\vec{u } \cdot \bm{\nabla} \mean{\Uvec}  -\alpha \theta \vec{g} - \bm{\nabla} \psi \nonumber \\ &- 2\,  \vec{\Omega} \times \vec{u } + \nu \bm{\nabla}^2 \vec{u } - \bm{\nabla} \cdot(\vec{u }\otimes\vec{u } - \meanRR), \\
\frac{\partial \theta }{\partial t} = &- \mean{\Uvec}\! \cdot \! \bm{\nabla} \theta - \vec{u} \!\cdot\! (\bm{\nabla} \mean{\Theta} + \bm{G}_0) + \chi \bm{\nabla}^2 \theta  \nonumber \!-\!\bm{\nabla} \!\cdot\! (\theta \vec{u }- \meanFF),
\end{align}
with $\otimes$ denoting the dyadic product
and $\bm{G}_0 = \nab T_0 - \vec{g}/c_p$.
For the Reynolds 
stress $\rij$, the turbulent heat flux $\meanF_i$ and the temperature variance $\meanQ$
we obtain
\begin{align}
\dot{\meanR}_{ij} &+ \mean{U}_k \pd_k \rij + \rik \pd_k \mean{U}_j + \rjk \pd_k \mean{U}_i + \alpha (\meanF_i g_j + \meanF_j g_i) \nonumber \\[.5mm]
 &-  \nu \pd_{kk} \rij + 2  \Omega_l\!\left(\varepsilon_{ilk}  \rjk +  \varepsilon_{jlk} \rik \right) \label{eq:er2}\\[.5mm] 
%&\hspace{-6.5mm}-\mean{u_i \pd_j \psi+u_j \pd_i \psi} - \mean{u_i \pd_k (u_j u_k)+u_j \pd_k (u_i u_k)} \\
= &-\mean{u_i \pd_j \psi+u_j \pd_i \psi} - \mean{u_i \pd_k (u_j u_k)+u_j \pd_k (u_i u_k)} \nonumber \\[.5mm]
%MJK Too wide
%-2 \nu\, \mean{\pd_k u_i \pd_k u_j},  \nonumber  \\[2mm]
&-2 \nu\, \mean{\pd_k u_i \pd_k u_j},  \nonumber  \\[2mm]
%\end{align}
%\begin{align}
 \dot{\meanF}_i &+ \mean{U}_j \pd_j \meanF_i + \meanF_j \pd_j \mean{U}_i + \rij (\pd_j \mean{\Theta}+ G_{0j})  + \alpha \meanQ g_i \nonumber \\[.5mm]  
 &- \onehalf (\nu + \chi) \pd_{jj} \meanF_i  + 2\varepsilon_{ijk}\Omega_j \meanF_k \label{eq:ef2}  \\[.5mm] 
= &-\mean{\theta \pd_i \psi}- \mean{\theta \pd_j (u_i u_j)+u_i \pd_j (\theta u_j)} \nonumber \\[.5mm]  
 & +\onehalf(\nu - \chi)\mean{\pd_k(\theta \pd_k u_i-u_i \pd_k \theta)}-(\nu + \chi)\mean{\pd_k\theta \partial_k u_i}, \nonumber \\[3mm]
 \dot{\meanQ} &+ \mean{U}_i \pd_i \meanQ + 2 \meanF_i (\pd_i \mean{\Theta}+  G_{0i}) - \chi \pd_{ii}\meanQ  \label{eq:eq2} \\
 =&-2 \mean{\theta \partial_i( \theta u_i)} 
 -2 \chi \mean{ (\partial_i \theta)^2}. \nonumber
\end{align}
Here,
the right hand sides 
contain third order correlations of 
fluctuating
quantities
(including the correlations with the pressure $\psi$)
and terms originating from the Laplacians which cannot be expressed by the considered second order correlations.
In the closure model of GOMS10 all these are replaced 
in the following way:
\begin{align}
&\hspace*{-0mm}\mean{u_i \pd_j \psi+u_j \pd_i \psi} + \mean{u_i \pd_k (u_j u_k)+u_j \pd_k (u_i u_k)} \\
&\hspace*{-0mm}+2 \nu \mean{\pd_k u_i \pd_k u_j} 
\rightarrow  \frac{C_1}{L} \meanR^{1/2} \rij + \frac{C_2}{L} \meanR^{1/2}(\rij - \onethird \meanR \delta_{ij}) \nonumber\\
& \hspace*{2.7cm}+ \nu \frac{C_\nu}{L^2} \rij  = \LamR \rij -\frac{C_2}{3L} \meanR^{3/2} \delta_{ij}, \nonumber
\end{align}
\begin{align}
&\hspace*{-2mm}\mean{\theta \pd_i \psi}+ \mean{\theta \pd_j (u_i u_j)+u_i \pd_j (\theta u_j)} \nonumber \\ 
& -\onehalf(\nu - \chi)\mean{\pd_k(\theta \pd_k u_i-u_i \pd_k \theta)}+(\nu + \chi)\mean{\pd_k\theta \partial_k u_i} \nonumber \\ 
& \rightarrow  \frac{C_6}{L} \meanR^{1/2} \meanF_i + \onehalf (\nu + \chi) \frac{C_{\nu\chi}}{L^2} \meanF_i = \LamF \meanF_i, \\[2mm]
&\hspace*{-2mm} 2 \big(\mean{\theta \partial_i (\theta u_i)} + \chi \mean{ (\partial_i \theta)^2}\big)\rightarrow \frac{C_7}{L} \meanR^{1/2}\! \meanQ \!+\! \chi \frac{C_{\chi}}{L^2} \meanQ = \LamQ \meanQ, \hspace*{-1mm}
\end{align}
with 
\begin{align}
  \LamR &= \frac{(C_1+C_2)}{L} \meanR^{1/2}  +\nu \frac{C_\nu}{L^2}, \\
    \LamF &=\frac{C_6}{L} \meanR^{1/2} + \onehalf (\nu+\chi) \frac{C_{\nu \chi}}{L^2}, \;
   \LamQ =   \frac{C_7}{L} \meanR^{1/2} + \chi \frac{C_{\chi}}{L^2}. \nonumber
\end{align}
Thus, the closure consists of relaxation terms, such as 
those $ \sim \rij$, isotropization terms 
$\sim (\rij - \onethird \meanR \delta_{ij})$ and terms like 
$\nu C_\nu L^{-2} \rij$ 
corresponding with
diffusion. 
For the length scale $L$, the distance to the closest 
boundary is adopted,
making the closure coefficients explicitly position dependent.
Applying 
the above ansatzes
we arrive at the equations 
\begin{align}
\dot{\meanR}_{ij} &+ \mean{U}_k \pd_k \rij + \rik \pd_k \mean{U}_j + \rjk \pd_k \mean{U}_i - \nu \pd_{kk} \rij   \nonumber \\[1mm] 
& + \alpha (\meanF_i g_j + \meanF_j g_i) + 2 \Omega_{l}\! \left(\varepsilon_{ilk} \rjk + \varepsilon_{jlk} \rik \right)\label{eq:clor2} \\ 
&=  -\left( \frac{C_1+C_2}{L} \meanR^{1/2} + \nu \frac{C_\nu}{L^2} \right)\rij + \frac{C_2}{3L} \meanR^{3/2} \delta_{ij} , \nonumber \\[3mm]
 \dot{\meanF}_i &+ \mean{U}_j \pd_j \meanF_i  + \meanF_j \pd_j \mean{U}_i + \rij (\pd_j \mean{\Theta}+  G_{0j}) + \alpha \meanQ g_i \nonumber \\[1mm]  
 &- \onehalf (\nu + \chi) \pd_{jj} \meanF_i  + 2\varepsilon_{ijk}\Omega_j \meanF_k \label{eq:clof2}\\ 
 &= - \left(\frac{C_6}{L} \meanR^{1/2}  + \onehalf (\nu + \chi) \frac{C_{\nu\chi}}{L^2}\right) \meanF_i, \nonumber \\[1mm] 
\dot{\meanQ} &+ \mean{U}_i \pd_i \meanQ + 2 \meanF_i (\pd_i \mean\Theta +  G_{0i} ) - \chi \pd_{ii}\meanQ  \nonumber \\
  &= -\!\!\left(\frac{C_7}{L} \meanR^{1/2}  + \chi \frac{C_{\chi}}{L^2} \right)\!\meanQ. \label{eq:cloq2}
\end{align}
Assuming now periodicity in the $x$ and $y$ directions we define the mean suitably as the average 
 over $x$ and $y$, $\mean{f}(z)\!=\!\int_{L_x} \! \int_{L_y} \!f(x,y,z) \,dx dy/L_x L_y$.
Ho\-ri\-zon\-tal derivatives vanish 
and
the continuity equation reduces to 
$\pd_z\mean{U}_z=0$,
hence
$\mean{U}_z=\mbox{const.}$   
For a plane layer with impenetrable boundaries this yields $\mean{U}_z=0$.
With gravity in 
 $z$ direction,
the equations 
for the remaining components of the mean velocity read
\EQ
\begin{aligned}
\dotm{U}_x &= -\pd_z \rxz + 2\Omz \mean{U}_y + \nu \pd_{zz}\mean{U}_x,\\
\dotm{U}_y &= -\pd_z \ryz - 2  \Omz \mean{U}_x + \nu \pd_{zz}\mean{U}_y.
\end{aligned} \label{eq:Uxy2}
\EE
Note that we do not need to solve for the mean 
reduced
pressure
 $\mean{\Psi}$ as it 
 does
 only affect $\mean{U}_z$. The 
equation for $\mean{\Theta}$ reduces to
\begin{equation}
\dotm{\Theta} = \chi \pd_{zz} \mean{\Theta} - \pd_z \meanF_z. \label{eq:thet2}
\end{equation} 
For Eqs.~\eqref{eq:clor2}--\eqref{eq:cloq2} we have now (with $\meanU_z=0$, $g=g_z$)
\begin{alignat}{2}
\dot{\meanR}_{xx} &=&& -2 \rxz \pd_z \mean{U}_x + 4 \Omz \rxy + \nu \pd_{zz} \rxx \nonumber \\ &&& -\LamR \rxx + \frac{C_2}{3L} \meanR^{3/2} ,\label{comprxx} \\
\dot{\meanR}_{xy} &=&& -\rxz \pd_z \mean{U}_y -\ryz \pd_z \mean{U}_x + 2 \Omx \rxz  \\ &&&+ 2 \Omz (\ryy-\rxx) + \nu \pd_{zz} \rxy -\LamR \rxy, \nonumber \\[1mm]
\dot{\meanR}_{xz} &=&& -\rzz \pd_z \mean{U}_x -\alpha \meanF_x g - 2 \Omx \rxy + 2 \Omz \ryz  \nonumber \\ &&&+ \nu \pd_{zz} \rxz  -\LamR \rxz, \\[1mm]
%\end{alignat}
%\begin{alignat}{1}
\dot{\meanR}_{yy} &=&& -2 \ryz \pd_z \mean{U}_y + 4 \Omx \ryz - 4 \Omz \rxy + \nu \pd_{zz} \ryy \nonumber \\ &&& - \LamR \ryy + \frac{C_2}{3L} \meanR^{3/2} , \\
\dot{\meanR}_{yz} &=&& -\rzz \pd_z \mean{U}_y -\alpha \meanF_y g + 2 \Omx (\rzz-\ryy) \nonumber \\ &&& - 2 \Omz \rxz   + \nu \pd_{zz} \ryz - \LamR \ryz, \\
\dot{\meanR}_{zz} &=&&  -2\alpha \meanF_z g - 4 \Omx \ryz + \nu \pd_{zz} \rzz -\LamR \rzz \nonumber \\ &&& + \frac{C_2}{3L} \meanR^{3/2}\!,\label{comprzz} \\[1mm]
%\end{eqnarray}
%\begin{eqnarray}
\dot{\meanF}_x &=&& -\rxz (\pd_z \mean{\Theta} + G_0) - \meanF_z \pd_z \mean{U}_x + \onehalf (\nu+\chi) \pd_{zz}\meanF_x \nonumber \\ &&& + 2 \Omz \meanF_y - \LamF \meanF_x, \\[2mm]
\dot{\meanF}_y &=&& -\ryz (\pd_z \mean{\Theta} + G_0) - \meanF_z \pd_z \mean{U}_y + \onehalf (\nu+\chi) \pd_{zz}\meanF_y \nonumber \\ &&& + 2 \Omx \meanF_z  - 2 \Omz \meanF_x - \LamF \meanF_y, \\[2mm]
\dot{\meanF}_z &=&&  -\rzz (\pd_z \mean{\Theta}  + G_0)- \alpha \meanQ g + \onehalf (\nu+\chi) \pd_{zz} \meanF_z \nonumber \\ &&&- 2 \Omx\meanF_y - \LamF \meanF_z,\\[2mm]
\dot{\meanQ} &=&& - 2 \meanF_z (\pd_z \mean{\Theta} + G_0) + \chi \pd_{zz} \meanQ - \LamQ \meanQ \, . \label{compQ}
\end{alignat}
with $G_0=G_{0z}$. Note that the stationary version of the autonomous system
\eqref{eq:Uxy2}--\eqref{compQ} 
does have non-trivial solutions as demonstrated in GOMS10. Due to the
nonlinearity of the system they exist not only for specific
combinations of its parameters like in linear eigenvalue problems, but
(at least within wide margins) for any specification of them.

For $\Omega_x=0$, that is, at the pole, there is a special stationary solution of the system \eqref{eq:Uxy2}--\eqref{compQ} 
characterized by
$\meanU_{x,y} = \meanR_{xy,xz,yz}=\meanF_{x,y}=0,\; \meanR_{xx}=\meanR_{yy}$ which is 
not explicitly dependent on
$\Omega_z$ and hence identical
with the corresponding solution of the non-rotating case.

\section{Closure model equations for the homogeneous Boussinesq system}
\label{app:eq0}

In this case we redefine
the average as a volume rather than a horizontal one, making the mean quantities indepedent of $z$, and obtain
\begin{align}
&\hspace*{-3.5mm}\begin{aligned}
\dot{\meanR}_{xx} &= 4 \Omz \rxy -\LamR \rxx  + \frac{C_2}{3L} \meanR^{3/2} , \\
\dot{\meanR}_{xy} &= 2 \Omx \rxz + 2 \Omz (\ryy-\rxx)  -\LamR \rxy, \\
\dot{\meanR}_{xz} &=  -\alpha \meanF_x g - 2 \Omx \rxy + 2 \Omz \ryz - \LamR \rxz,  \\
\dot{\meanR}_{yy} &= 4 \Omx \ryz - 4 \Omz \rxy -\LamR \ryy  + \frac{C_2}{3L} \meanR^{3/2} ,  \label{eq:zdclo}  \\
\dot{\meanR}_{yz} &= -\alpha \meanF_y g + 2 \Omx (\rzz-\ryy) - 2 \Omz \rxz \!-\! \LamR \ryz, \\
\dot{\meanR}_{zz} &=  -2\alpha \meanF_z g - 4 \Omx \ryz - \LamR \rzz  + \frac{C_2}{3L} \meanR^{3/2}, 
\end{aligned} \hspace{-1cm} \\
&\hspace*{-1.5mm}\begin{aligned}
\dot{\meanF}_x &= -\rxz G_0 + 2 \Omz \meanF_y  - \LamF \meanF_x,\\
\dot{\meanF}_y &= -\ryz G_0 + 2 \Omx \meanF_z - 2 \Omz \meanF_x - \LamF \meanF_y, \\
\dot{\meanF}_z &=  -\rzz G_0  - \alpha \meanQ g - 2 \Omx \meanF_y - \LamF \meanF_z, \\[1mm]
\dot{\meanQ}    &=- 2 \meanF_z G_0 - \LamQ \meanQ, 
\end{aligned}
\end{align}
which is for $\vec{\Omega}=\vec{0}$ equivalent to Eqs. (53) of GOMS10.
The resulting equation for $\meanR$ reads
\EQ
  \dot{\meanR} =  -2\alpha \meanF_z g - \frac{C_1}{L} \meanR^{3/2}
\EE
and is not explicitly influenced by rotation.

In the non-rotating case the equations for $\dot{\meanR}_{xx}$, $\dot{\meanR}_{yy}$, $\dot{\meanR}_{zz}$, $\dot{\meanF}_z$ and $\dot{\meanQ}$
form a closed system which can be solved in separation from the remaining equations. Once the solution of the former is known
the latter can be solved where one finds again two separate systems: $\{\dot{\meanR}_{xz},\dot{\meanF}_x\}$ and $\{\dot{\meanR}_{yz},\dot{\meanF}_y\}$.
They
have the same shape, and when assuming that there is a stationary solution for $\meanR$ from the first system, we arrive at  
the eigenvalue problem for the growth rate of an ansatz $\meanR_{xz},\meanF_x,\meanR_{yz},\meanF_y \sim \exp(\lambda t)$
\EQ
\left| \begin{matrix}
             \LamR+\lambda  &   \alpha g \\
                G_0                    &   \LamF+\lambda
        \end{matrix}
\right| =0
\EE
with constant $\LamR,\LamF$. The solutions are
\EQ
    \lambda_{1,2} = -\frac{\LamR+ \LamF}{2} \pm \sqrt{\frac{(\LamR - \LamF)^2}{4} + \alpha g G_0}     \label{eq:lams}
\EE
and given that $\alpha g G_0 > 0$ for convection, unstable solutions cannot completely be ruled out for sufficiently large values of this product,
but 
had
most likely to be considered unphysical.

Nontrivial closed form {\em stationary} solutions can be derived for the special settings $\ve{\Omega}=\ve{0}$ or $\vartheta=0$ (pole):
In both cases we have as in the inhomogeneous case
\begin{align}
\meanR_{xy,xz,yz}&=\meanF_{x,y}=0, \quad\meanR_{xx,yy} = \frac{C_2}{3(C_1+C_2)} \meanR  \label{eq:specpole1}
\end{align}
{hence}
\begin{align}
 \meanR&= \frac{2 \alpha g G_0 L^2}{C_1 C_6} \left( \frac{C_1}{C_7} + \frac{3 C_1 + C_2}{3(C_1+C_2)} \right) \label{eq:meanR} \\
  \meanR_{zz}&=  \frac{3 C_1 + C_2}{3(C_1+C_2)} \meanR, \quad
 \meanF_z = -\frac{C_1}{2\alpha g L} \meanR^{3/2} \label{eq:meanF}\\
 \meanQ &= \frac{G_0}{\alpha g}\frac{C_1}{C_7} \meanR \label{eq:specpole2}
\end{align}
which coincides with the solution given in GOMS10.
In turn it is under these conditions possible to determine the $\{C_i\}$ uniquely when $\meanR_{xx}=\meanR_{yy}$, $\meanR_{zz}$, $\meanF_z$ and $\meanQ$
are given from a DNS:
\EQ
\begin{aligned}
3 \meanR_{xx} C_1  &- \left(\meanR_{zz}-\meanR_{xx} \right)C_2=0 \\
3 \meanR_{zz} C_1 &+ (3 \meanR_{zz} - \meanR) C_2 = - 6\alpha g L \meanF_z / \meanR^{1/2} \\
C_6 &= - (G_0 \meanR_{zz} + \alpha g \meanQ) L /\meanF_z  \meanR^{1/2}  \\
C_7 &= -2 G_0 L \meanF_z / \meanQ \meanR^{1/2} .
\end{aligned}  \label{eq:Csol}
\EE
Inserting \eqref{eq:meanR} in \eqref{eq:lams} we obtain
\begin{align}
     &(\LamR - \LamF)^2 + 4\alpha g G_0  - (\LamR + \LamF)^2 \label{eq:stabnorot} \\&= 4(\alpha g G_0 - \LamR\LamF )
     =2 \alpha g G_0 \left(1- \frac{C_1+C_2}{C_7} - \frac{C_2}{3C_1} \right) \nonumber
\end{align}
the sign of which depends solely on the parameters $\{C_i\}$ and not on $\alpha g G_0$. Requiring \eqref{eq:stabnorot} to be negative provides an additional constraint. 
A corresponding generalized condition, ensuring overall stability, is referred to in Sec.~\ref{sec:opt}.

Another special situation is found at the equator ($\vartheta=90\degr$, hence $\Omega_z=0$) where the system \eqref{eq:zdclo} decomposes into a closed one for 
the quantities $\rxx$, $\ryy$, $\ryz$, $\rzz$, $\meanF_y$, $\meanF_z$ and $\meanQ$ and another one for $\rxy$, $\rxz$ and $\meanF_x$ which can be solved
once $\meanR$ from the first system is known. The latter reads in the stationary case
\begin{align}
0&= 2 \Omx \rxz   -\LamR \rxy, \nonumber\\
0&=  -\alpha \meanF_x g - 2 \Omx \rxy - \LamR \rxz, \nonumber \\
0&= -\rxz G_0  - \LamF \meanF_x, \nonumber
\end{align}
%MR:
%with 
where
$\meanF_x$ 
%being 
can be
eliminated  by  the last line.
% by $\meanF_x = - G_0 \rxz /\LamF$.
The remaining two equations form a homogeneous linear system for 
$\rxy$ and $\rxz$ having the determinant
\[
   \alpha g G_0 \frac{C_1+C_2}{C_6} - \left(\frac{C_1+C_2}{L} \right)^2 \! \meanR - 4 \Omega_x^2.
\]
Nontrivial solutions would be possible if $\meanR$ were to assume a special value depending on the parameters. However, this has the unphysical
consequence of $\rxy$, $\rxz$ and $\meanF_x$ becoming dependent on an arbitrary quantity. So we have to conclude, that they either vanish or are
time-dependent. In the latter case we have to require stability, so these quantities 
were
bound to decay to zero or to perform stationary oscillations with an arbitrary amplitude. 
As the only physically meaningful option we assume that they vanish.
The remaining system reads
\begin{align}
0 &=  -\LamR \rxx  + \frac{C_2}{3L} \meanR^{3/2} , \nonumber\\
0 &= 4 \Omx \ryz  -\LamR \ryy  + \frac{C_2}{3L} \meanR^{3/2} , \nonumber \\
0 &= -\alpha \meanF_y g + 2 \Omx (\rzz-\ryy)  - \LamR \ryz, \nonumber \\
0 &=  -2\alpha \meanF_z g - 4 \Omx \ryz - \LamR \rzz  + \frac{C_2}{3L} \meanR^{3/2},\nonumber \\[1mm]
0 &= -\ryz G_0 + 2 \Omx \meanF_z  - \LamF \meanF_y, \nonumber\\[1mm]
0 &=  -\rzz G_0  - \alpha \meanQ g - 2 \Omx \meanF_y - \LamF \meanF_z, \nonumber\\[1mm]
0 &= - 2 \meanF_z G_0 - \LamQ \meanQ. \nonumber
\end{align}
 From the first line it follows $\meanR = C_{12}(\meanR_{yy}+\meanR_{zz}), \, C_{12} = 3(C_1+C_2)/(3 C_1 + 2 C_2)$, from the last $\meanQ = -2 G_0 \meanF_z / \LamQ$ leaving a system with five variables
 only. It can be broken down to a nonlinear equation for $\meanR$ which is (apart from $\meanR=0$) solved by  the solutions of
\begin{equation}
   2\Omega_x\big(2K - \frac{\meanR E(\meanR)}{C_{12}}\big) - \LamR K D(\meanR)  - \alpha g G(\meanR) E(\meanR) = 0, \label{eq:equsol}
\end{equation}
 completed by 
 \begin{align*}
 %MR: simplified
% \meanF_z &= K \meanR^{3/2}, \quad \meanF_y = G(\meanR)  \meanR^{3/2}, \\
 \meanF_z &= K \meanR^{3/2}, \quad \meanF_y = G(\meanR)  \meanR^{3/2}, \quad  K=-C_1/L,\\
 \meanR_{yy} &= \frac{\meanR}{C_{12}} - \frac{K}{E(\meanR)} \meanR^{3/2}, \quad \meanR_{yz} = \frac{K D(\meanR)}{E(\meanR)} \meanR^{3/2}, \\
 \meanR_{zz} &= \frac{K}{E(\meanR)} \meanR^{3/2},      %\quad K = \frac{2 C_2}{3L}  - \frac{C_1 + C_2}{C_{12} L }, \quad
 \end{align*}
 with 
 \begin{align*}
 %MR: arguments removed/added
    E(\meanR) &= A_z D + B_z, \; G(\meanR) = \frac{1}{K} \frac{A_z D + B_z}{A_y D + B_y}\\[1mm]
  D(\meanR) &= \alpha g \frac{4 \Omega_x B_z + \LamR B_y}{16 \Omega_x^2 + \LamR^2 - \alpha g \big(4\Omega_x A_z -\LamR A_y\big)}\\[1mm]
  A_y(\meanR) &= \frac{G_0}{\Delta} (\alpha g G_0/\LamQ -\LamF), \\
   B_y(\meanR) &= -G_0 2\Omega_x/\Delta = -A_z, \;
   B_z(\meanR) = - G_0 \LamF/\Delta, \\
  \Delta(\meanR) &= 4 \Omega_x^2 + \LamF(\LamF - \alpha g 2 G_0/\LamQ).
%    E(\meanR) &= A_z(\meanR) D(\meanR) + B_z(\meanR), \; G(\meanR) = \frac{1}{K} \frac{A_z D + B_z}{A_y D + B_y}(\meanR)\\[1mm]
%  D(\meanR) &= \alpha g \frac{4 \Omega_x B_z(\meanR) + \LamR B_y(\meanR)}{16 \Omega_x^2 + \LamR^2 - \alpha g \big(4\Omega_x A_z(\meanR) -\LamR A_y(\meanR)\big)}\\[1mm]
%  A_y &= \frac{G_0}{\Delta} (\alpha g G_0/\LamQ -\LamF), \; B_y = -G_0 2\Omega_x/\Delta = -A_z, \\
%   B_z &= - G_0 \LamF/\Delta, \quad
%  \Delta = 4 \Omega_x^2 + \LamF(\LamF - \alpha g 2 G_0/\LamQ).
 \end{align*}
It cannot be guaranteed that \eqref{eq:equsol} has positive solutions for $\meanR$ for any arbitrary set of parameters, in particular for arbitrary positive $\{C_i\}$.

In contrast, for $\ve{\Omega}\ne\ve{0}$ and $\vartheta\ne0,90^\circ$ none of the components of $\meanRR$  and $\meanFF$ disappear and the determination of the $\{C_i\}$ from
DNS results has to deal with an overdetermined system: ten equations vs. four unknowns.

With respect to the realizability constraint \eqref{eq:stcon}, an
analysis analogous to that of GOMS10, App.~A, but with rotation
included, leads to the following relation for the temporal derivative
of
the quantity $\meanT = X_iT_{ij}X_j = X_i(\meanR_{ij} - \meanQ^{-1} \meanF_i \meanF_j ) X_j$
\begin{align}
\partial_t {\meancT}  =  \,& (2C_6 -C_7 - C_1 - C_2) (\meanFF \cdot \vec{X})^2 \frac{ \sqrt{\meanR}}{ \meanQ L}\nonumber \\
& + \frac{C_2}{3 L} \meanR^{3/2} \vec{X}^2 - ( C_1+ C_2) \frac{\meancT \sqrt{\meanR}}{L}  \label{equ:realo} \\ 
& + 2 (\meanFF \cdot \vec{X}) \meanQ^{-1} X_j T_{jz}  G_0   - 4 X_i \varepsilon_{ilk} \Omega_l T_{jk} X_j . \nonumber
\end{align}
Repeating the arguments of GOMS10 here, one finds that the realizability condition is not affected by the presence of rotation,
since $\Omega_l$ in \eqref{equ:realo} is multiplied by the vanishing term  $T_{ij}X_j$.
Similarly, by retaining the model coefficients $C_{\nu}$, $C_{\kappa}$ and 
$C_{\nu\kappa}$ one can derive the following expression
\begin{align*}
\partial_t {\meancT}  = &
\bigg( \!2C_6 -C_7 - C_1 - C_2 \\
&\:+ \frac{2 (\nu + \kappa)C_{\nu\kappa}-\kappa C_{\kappa}-\nu C_{\nu}}{L\sqrt{\meanR}} \!\bigg) 
(\meanFF \cdot \vec{X})^2\frac{ \sqrt{\meanR}}{\meanQ L}  \\
&+ \frac{C_2}{3 L} \meanR^{3/2} \vec{X}^2 - ( C_1+ C_2) \frac{\meancT \sqrt{\meanR}}{L}  \\
& + 2 (\meanFF \cdot \vec{X}) \meanQ^{-1} X_j T_{jz} G_0
  - 4 X_i \varepsilon_{ilk} \Omega_l T_{jk} X_j  \\ 
& - X_i \frac{\nu C_{\nu}}{L^2} T_{ij} X_j \,,
\end{align*}
from which one obtains the 
realizability
criterion 
\begin{equation}
2C_6 -C_7 - C_1 - C_2 + \frac{2 (\nu + \kappa)C_{\nu\kappa}-\kappa C_{\kappa}-\nu C_{\nu}}{L\sqrt{\meanR}} \ge 0.
%MR:
%\label{equ:realnk}
\nonumber
\end{equation}
This criterion cannot be formulated as a condition for the model
parameters alone,
unlike \eqref{eq:stcon}.
However, we can infer the two {\em sufficient} conditions \eqref{eq:stcon} and $2 (\nu + \kappa)C_{\nu\kappa}-\kappa C_{\kappa}-\nu C_{\nu}\ge0$.
With $\Pra=1$ the latter one can be written as 
$4 C_{\nu\kappa}- C_{\kappa}- C_{\nu}\ge0$, which is satisfied by the values 
$C_{\nu\kappa} \approx 6$, $C_{\nu} \approx 12$ and $C_{\kappa} \approx 2$ 
given in GOMS10
and also by our result \eqref{eq:unival}.

\begin{table*}
  \centering
  \caption[]{Summary of the Boussinesq DNS results.  Normalizations  (indicated by a tilde) are carried out 
with $\alpha g d^2 G_0$ for Reynolds stress,  with $d^2 G_0^{3/2} (\alpha g)^{1/2}$ for heat flux,
and with $(d G_0)^2$ for temperature variance.
The grid size in all runs in Sets~Z and A--G is $64^3$. 
We have reproduced a subset of these runs with $128^3$ grid and confimed
that the results are typically within ten per cent of the lower resolution 
ones. However, see Sect.~\ref{sec:DNSruns} for the convergence issue 
related to the time step.
}
  \vspace{-0.5cm}
  \label{tab:bres}
  \arraycolsep5pt
  $$
  \begin{array}{@{\hspace*{0mm}}
%1                           2                                    3                                       4                                      5                                    6                                   7                                   8                                   9                                10                                 11                                                                      %                                12                                     13                            14				 15
  p{0.02\linewidth} r @{\hspace*{1.5mm}}r @{\hspace*{1.5mm}}r @{\hspace*{1.5mm}} r @{\hspace*{3.mm}}r @{\hspace*{2.mm}}r @{\hspace*{2.5mm}}r @{\hspace*{4mm}}r @{\hspace{1.mm}}r @{\hspace*{3.mm}}r @{\hspace*{1.5mm}}r   @{\hspace*{2.5mm}}r  @{\hspace*{2mm}} r @{\hspace*{2.mm}} r @{\hspace*{0mm}}}
    \hline
    \noalign{\smallskip}
%MJK Too wide
%    Run &  \vartheta\: &\Tay/10^{6} & \Co\; & \Rey & \trxx\; & \trxy/10^{-2}  & \trxz/10^{-2} & \tryy\; & \tryz/10^{-2} & \trzz\; & \tilde{\meanF}_x/10^{-2} & \tilde{\meanF}_y/10^{-2} & \tilde{\meanF}_z\; & \tilde{\meanQ}\;\;\: \\
    Run &  \vartheta\: &\Tay \; & \Co\; & \Rey & \trxx\; & \trxy\; & \trxz \; & \tryy\; & \tryz \; & \trzz\; & \tilde{\meanF}_x \; & \tilde{\meanF}_y \; & \tilde{\meanF}_z\; & \tilde{\meanQ}\;\;\: \\
    &&[10^{6}] &&&&[10^{-2}] &[10^{-2}] &&[10^{-2}] &&[10^{-2}] &[10^{-2}] &&\\
    \noalign{\smallskip}
    \hline\hline\\[-3mm]
Z &0\degr &0.00 &0.00 &          87 &0.114 &0.083 &-0.134 &0.113 &0.139 &0.365 &-0.169 &0.105 &0.286 &0.333 \\[.5mm]
\hline\\[-3mm]
A1 &0\degr &0.04 &0.06 &          91 &0.119 &-0.013 &0.025 &0.119 &-0.145 &0.412 &-0.025 &-0.178 &0.329 &0.381 \\
A2 &0\degr &0.16 &0.11 &          92 &0.123 &-0.030 &-0.117 &0.123 &-0.088 &0.419 &-0.166 &-0.095 &0.333 &0.386 \\
A3 &0\degr &0.36 &0.16 &          95 &0.127 &0.009 &0.006 &0.128 &-0.015 &0.455 &0.051 &-0.098 &0.362 &0.417 \\
A4 &0\degr &1.00 &0.26 &          97 &0.131 &-0.021 &0.022 &0.131 &-0.043 &0.481 &0.049 &-0.040 &0.382 &0.436 \\
A5 &0\degr &1.44 &0.31 &         100 &0.135 &0.005 &0.049 &0.136 &0.068 &0.511 &0.047 &0.044 &0.406 &0.462 \\
A6 &0\degr &2.56 &0.39 &         103 &0.143 &0.032 &0.070 &0.144 &-0.142 &0.548 &0.128 &-0.191 &0.428 &0.480 \\
A7 &0\degr &12.96 &0.76 &         120 &0.174 &0.057 &-0.670 &0.175 &0.053 &0.795 &-0.614 &0.048 &0.611 &0.656 \\
A8 &0\degr &100 &1.71 &         148 &0.221 &-0.132 &-0.373 &0.219 &-0.524 &1.285 &-0.417 &-0.379 &0.984 &1.021 \\
A9 &0\degr &400 &3.29 &         154 &0.225 &-0.115 &-0.609 &0.217 &-0.629 &1.424 &-0.735 &-0.564 &1.067 &1.096 \\
A10 &0\degr &10000 &10.79 &         235 &0.252 &0.016 &0.295 &0.254 &-0.582 &3.849 &-0.005 &-0.241 &2.947 &2.839 \\[.5mm]
\hline\\[-3mm]
B1 &15\degr &0.04 &0.06 &          90 &0.121 &-0.045 &-0.275 &0.120 &-1.211 &0.400 &-0.220 &-1.301 &0.316 &0.367 \\
B2 &15\degr &0.16 &0.11 &          90 &0.118 &-0.012 &-0.173 &0.121 &-2.358 &0.395 &-0.188 &-2.464 &0.312 &0.363 \\
B3 &15\degr &0.36 &0.18 &          85 &0.110 &0.098 &-0.504 &0.117 &-3.191 &0.350 &-0.650 &-3.384 &0.276 &0.326 \\
B4 &15\degr &1.00 &0.30 &          85 &0.110 &0.213 &-0.947 &0.122 &-4.456 &0.337 &-1.152 &-4.955 &0.268 &0.319 \\
B5 &15\degr &1.44 &0.36 &          84 &0.109 &0.313 &-1.106 &0.123 &-4.479 &0.323 &-1.459 &-5.120 &0.257 &0.310 \\
B6 &15\degr &2.56 &0.51 &          80 &0.104 &0.316 &-1.313 &0.119 &-4.158 &0.284 &-1.875 &-4.963 &0.227 &0.279 \\
B7 &15\degr &5.76 &0.75 &          81 &0.110 &0.190 &-1.613 &0.125 &-3.797 &0.278 &-2.330 &-5.123 &0.223 &0.279 \\
B8 &15\degr &10.24 &0.99 &          82 &0.119 &-0.021 &-1.421 &0.127 &-3.261 &0.286 &-2.112 &-4.939 &0.226 &0.284 \\
B9 &15\degr &16 &1.27 &          80 &0.119 &-0.030 &-0.980 &0.125 &-2.528 &0.261 &-1.514 &-4.997 &0.200 &0.256 \\
B10 &15\degr &100 &1.73 &         146 &0.447 &1.485 &29.507 &0.258 &-4.673 &0.987 &18.614 &-10.961 &0.657 &0.760 \\
B11 &15\degr &400 &1.71 &         296 &1.988 &3.467 &201.160 &0.615 &-9.467 &4.323 &123.721 &-23.260 &2.725 &2.650 \\[.5mm]
\hline\\[-3mm]
C1 &30\degr &0.04 &0.06 &          89 &0.116 &0.022 &-0.103 &0.119 &-2.333 &0.390 &-0.175 &-2.492 &0.310 &0.362 \\
C2 &30\degr &0.16 &0.12 &          86 &0.110 &0.093 &-0.470 &0.120 &-3.776 &0.352 &-0.531 &-4.210 &0.283 &0.336 \\
C3 &30\degr &0.36 &0.19 &          80 &0.099 &0.277 &-0.587 &0.114 &-4.365 &0.289 &-0.787 &-5.068 &0.234 &0.286 \\
C4 &30\degr &1.00 &0.34 &          74 &0.091 &0.561 &-1.261 &0.110 &-4.293 &0.232 &-1.762 &-5.499 &0.191 &0.244 \\
C5 &30\degr &1.44 &0.42 &          72 &0.088 &0.597 &-1.430 &0.108 &-3.892 &0.214 &-2.089 &-5.415 &0.177 &0.230 \\
C6 &30\degr &2.56 &0.58 &          70 &0.088 &0.531 &-1.565 &0.104 &-3.263 &0.200 &-2.350 &-5.005 &0.163 &0.217 \\
C7 &30\degr &5.76 &0.84 &          73 &0.098 &0.455 &-1.177 &0.108 &-2.732 &0.212 &-2.413 &-5.026 &0.168 &0.228 \\
C8 &30\degr &10.24 &1.02 &          80 &0.118 &0.362 &0.377 &0.123 &-2.702 &0.262 &-1.622 &-5.732 &0.196 &0.266 \\
C9 &30\degr &4.00 &0.70 &          73 &0.099 &0.417 &-1.223 &0.112 &-3.191 &0.208 &-2.463 &-5.620 &0.176 &0.249 \\
C10 &30\degr &16 &1.13 &          89 &0.157 &0.533 &3.083 &0.143 &-2.810 &0.331 &-0.544 &-6.472 &0.240 &0.369 \\
C11 &30\degr &100 &1.14 &         223 &1.023 &2.198 &83.326 &0.483 &-7.129 &2.419 &28.885 &-22.541 &1.501 &2.178 \\
C12 &30\degr &400 &1.24 &         407 &3.727 &5.181 &297.646 &1.053 &-12.171 &8.305 &26.867 &-52.942 &5.341 &11.037 \\
C13 &30\degr &10000 &3.18 &         798 &13.330 &8.958 &722.547 &2.298 &-4.884 &34.601 &-717.873 &-85.046 &25.270 &173.730 \\[.5mm]
\hline\\[-3mm]
D1 &45\degr &0.04 &0.06 &          89 &0.117 &-0.021 &0.063 &0.122 &-3.071 &0.389 &0.054 &-3.341 &0.311 &0.365 \\
D2 &45\degr &0.16 &0.12 &          82 &0.101 &0.157 &-0.449 &0.119 &-4.722 &0.305 &-0.527 &-5.426 &0.247 &0.301 \\
D3 &45\degr &0.36 &0.20 &          75 &0.089 &0.405 &-0.635 &0.113 &-4.609 &0.239 &-0.885 &-5.861 &0.196 &0.249 \\
D4 &45\degr &1.00 &0.39 &          65 &0.075 &0.705 &-1.168 &0.097 &-3.566 &0.165 &-1.822 &-5.472 &0.138 &0.187 \\
D5 &45\degr &1.44 &0.47 &          64 &0.074 &0.755 &-1.218 &0.094 &-3.145 &0.158 &-1.982 &-5.180 &0.130 &0.180 \\
D6 &45\degr &2.56 &0.65 &          62 &0.075 &0.735 &-1.050 &0.087 &-2.470 &0.144 &-2.061 &-4.689 &0.116 &0.166 \\
D7 &45\degr &5.76 &0.83 &          74 &0.104 &0.701 &0.183 &0.110 &-2.450 &0.215 &-1.415 &-5.606 &0.160 &0.228 \\
D8 &45\degr &10.24 &0.89 &          91 &0.152 &0.816 &2.599 &0.152 &-2.936 &0.348 &-0.114 &-7.675 &0.247 &0.343 \\
D9 &45\degr &4.00 &0.72 &          70 &0.098 &0.527 &-0.058 &0.105 &-2.473 &0.188 &-1.131 &-5.279 &0.150 &0.234 \\
D10 &45\degr &16 &0.95 &         107 &0.218 &0.937 &8.224 &0.184 &-3.041 &0.505 &2.167 &-9.218 &0.354 &0.554 \\
D11 &45\degr &100 &1.02 &         247 &1.079 &1.361 &90.448 &0.535 &-5.660 &3.209 &24.256 &-27.786 &1.906 &2.882 \\
D12 &45\degr &400 &1.17 &         432 &3.277 &1.797 &300.328 &0.994 &-9.638 &10.468 &-25.959 &-59.461 &5.819 &13.090 \\
D13 &45\degr &10000 &2.22 &        1143 &24.515 &6.300 &1403.346 &5.534 &1.043 &73.121 &-4214.031 &-253.095 &76.100 &715.106 \\[.5mm]
\hline%\\[-2.5mm]
   \end{array}
   $$ 
\end{table*}
%}

%\onltab{2}{
\begin{table*}
  \centering
  \caption[]{Summary of the   
  DNS results continued.
  For normalizations  see Table\ref{tab:bres}.
}
  \vspace{-0.5cm}
  \label{tab:bres2}
  \arraycolsep5pt
  $$
%  \begin{array}{p{0.03\linewidth}ccccccccccccccccccccc}
%  \begin{array}{p{0.03\linewidth}rrrrrrrrrrrrrrrrrrrrr}
  \begin{array}{
%1                           2                                 3  45                                  6                                     7                                   8                                   9                                10                                 11                               12 13 14				 15
  p{0.02\linewidth} r @{\hspace*{2mm}}r r  r @{\hspace*{3.5mm}}r @{\hspace*{3.mm}}r @{\hspace*{3mm}}r @{\hspace*{4mm}}r @{\hspace{3mm}}r @{\hspace*{4mm}}r @{\hspace*{3mm}}r   r  @{\hspace*{4mm}}   r @{\hspace*{3.5mm}} r}
    \hline
    \noalign{\smallskip}
%     Run &  \vartheta  & \Tay/10^{6} & \Co & \Rey & \trxx & \trxy/10^{-2}  & \trxz/10^{-2} & \tryy & \tryz/10^{-2} & \trzz & \tilde{\meanF}_x/10^{-2} & \tilde{\meanF}_y/10^{-2} & \tilde{\meanF}_z & \tilde{\meanQ} \\[.5mm]
%MJK Too wide
%    Run &  \vartheta\: &\Tay/10^{6} & \Co\; & \Rey & \trxx\; & \trxy/10^{-2}  & \trxz/10^{-2} & \tryy\; & \tryz/10^{-2} & \trzz\; & \tilde{\meanF}_x/10^{-2} & \tilde{\meanF}_y/10^{-2} & \tilde{\meanF}_z\; & \tilde{\meanQ}\;\;\: \\
    Run &  \vartheta\: &\Tay & \Co\; & \Rey & \trxx\; & \trxy & \trxz& \tryy\; & \tryz& \trzz\; & \tilde{\meanF}_x& \tilde{\meanF}_y & \tilde{\meanF}_z\; & \tilde{\meanQ}\;\;\: \\
    &&[10^{6}] &&&&[10^{-2}] &[10^{-2}] &&[10^{-2}] &&[10^{-2}] &[10^{-2}] &&\\
    \noalign{\smallskip}
        \hline\hline\\[-3mm]
E1 &60\degr &0.04 &0.06 &          87 &0.112 &0.034 &-0.115 &0.120 &-3.451 &0.362 &-0.107 &-3.774 &0.288 &0.339 \\
E2 &60\degr &0.16 &0.13 &          76 &0.091 &0.180 &-0.280 &0.113 &-4.469 &0.253 &-0.398 &-5.385 &0.205 &0.254 \\
E3 &60\degr &0.36 &0.22 &          69 &0.078 &0.437 &-0.514 &0.106 &-4.202 &0.192 &-0.792 &-5.826 &0.159 &0.208 \\
E4 &60\degr &1.00 &0.43 &          59 &0.063 &0.640 &-0.714 &0.089 &-2.796 &0.128 &-1.252 &-5.021 &0.105 &0.151 \\
E5 &60\degr &1.44 &0.53 &          58 &0.062 &0.697 &-0.670 &0.081 &-2.293 &0.118 &-1.313 &-4.488 &0.096 &0.139 \\
E6 &60\degr &2.56 &0.68 &          60 &0.068 &0.721 &-0.379 &0.085 &-1.930 &0.128 &-1.087 &-4.602 &0.096 &0.146 \\
E7 &60\degr &5.76 &0.84 &          72 &0.099 &0.862 &0.070 &0.112 &-1.677 &0.204 &-0.658 &-5.227 &0.143 &0.213 \\
E8 &60\degr &10.24 &0.92 &          89 &0.138 &0.929 &0.437 &0.155 &-1.790 &0.326 &0.112 &-6.615 &0.219 &0.317 \\
E9 &60\degr &16.00 &1.01 &         101 &0.182 &1.256 &1.335 &0.181 &-1.271 &0.436 &0.261 &-6.841 &0.278 &0.386 \\[.5mm]
\hline\\[-3mm]
F1 &75\degr &0.04 &0.06 &          85 &0.108 &0.015 &0.001 &0.120 &-3.963 &0.347 &-0.028 &-4.396 &0.278 &0.331 \\
F2 &75\degr &0.16 &0.13 &          75 &0.089 &0.089 &-0.175 &0.114 &-4.675 &0.246 &-0.187 &-5.806 &0.201 &0.253 \\
F3 &75\degr &0.36 &0.23 &          66 &0.071 &0.287 &-0.332 &0.104 &-4.106 &0.171 &-0.493 &-5.951 &0.143 &0.193 \\
F4 &75\degr &1.00 &0.44 &          57 &0.055 &0.467 &-0.366 &0.086 &-2.566 &0.115 &-0.674 &-5.068 &0.094 &0.140 \\
F5 &75\degr &1.44 &0.56 &          54 &0.051 &0.544 &-0.327 &0.078 &-1.966 &0.100 &-0.659 &-4.454 &0.079 &0.123 \\
F6 &75\degr &2.56 &0.71 &          57 &0.059 &0.697 &-0.328 &0.083 &-1.418 &0.113 &-0.550 &-3.953 &0.083 &0.128 \\
F7 &75\degr &5.76 &0.91 &          67 &0.082 &1.364 &-0.582 &0.108 &-0.836 &0.162 &-0.572 &-3.556 &0.103 &0.138 \\
F8 &75\degr &12.96 &1.06 &          86 &0.134 &1.513 &-0.810 &0.158 &-0.715 &0.296 &0.201 &-4.813 &0.177 &0.275 \\[.5mm]
\hline\\[-3mm]
G1 &90\degr &0.04 &0.06 &          85 &0.108 &0.015 &-0.025 &0.120 &-3.925 &0.347 &-0.046 &-4.319 &0.279 &0.332 \\
G2 &90\degr &0.16 &0.14 &          75 &0.087 &0.010 &-0.064 &0.114 &-4.725 &0.240 &-0.073 &-5.925 &0.197 &0.250 \\
G3 &90\degr &0.36 &0.23 &          66 &0.069 &-0.012 &0.004 &0.107 &-4.144 &0.168 &-0.008 &-6.277 &0.141 &0.192 \\
G4 &90\degr &1.00 &0.44 &          58 &0.054 &0.096 &-0.121 &0.090 &-2.500 &0.120 &-0.265 &-5.076 &0.095 &0.141 \\
G5 &90\degr &1.44 &0.57 &          53 &0.045 &0.004 &0.018 &0.082 &-1.870 &0.098 &0.028 &-4.503 &0.075 &0.119 \\
G6 &90\degr &2.56 &0.86 &          47 &0.024 &0.065 &-0.031 &0.088 &-0.645 &0.062 &-0.049 &-2.702 &0.041 &0.075 \\
G7 &90\degr &4.00 &1.08 &          47 &0.018 &0.029 &0.026 &0.100 &-0.271 &0.058 &-0.077 &-1.572 &0.031 &0.060 \\[.5mm]
\hline
\hline
   \end{array}
   $$ 
\end{table*}
\begin{table*}
  \centering
  \caption[]{Summary of the   
  DNS results with different Rayleigh numbers. The non-primed runs are non-rotating, while
for the primed runs $\Tay=10^6$ and $\vartheta=0\degr$. For normalizations see Table~\ref{tab:bres}.
The grid resolutions are $64^3$ (R1--2), $128^3$ 
  (R3--4), $256^3$ (R5), and $512^3$ (R6).
}
  \vspace{-0.5cm}
  \label{tab:bresRa}
  \arraycolsep5pt
  $$
  \begin{array}{p{0.03\linewidth}rrrrccccccccccccccccc@{\hspace{*13mm}}c}
    \hline
    \noalign{\smallskip}
    Run &\Ray & \;\Rey & \;\Co & \trxx & \trxy& \trxz& \tryy & \tryz& \trzz & \tilde{\meanF}_x& \tilde{\meanF}_y& \tilde{\meanF}_z & \tilde{\meanQ} \\
    &[10^{6}] &&&&[10^{-2}] &[10^{-2}] &&[10^{-2}] &&[10^{-2}] &[10^{-2}] &&\\
    \noalign{\smallskip}
    \hline\hline\\[-3mm]
R1  &        0.25 &         \phm\phm 68 &\phm\phm0   & 0.126 &-0.029    &\phm0.018 &0.128 &\phm0.041 &0.480 &\phm0.064 &\phm0.063 &0.403 &0.498 \\
R2  &           1 &         \phm128     &\phm\phm0   & 0.123 &\phm0.032 &-0.209    &0.124 &\phm0.001 &0.395 &-0.331    &\phm0.015 &0.317 &0.411 \\
R3  &           4 &         \phm259     &\phm\phm0   & 0.131 &-0.014    &\phm0.115 &0.131 &-0.090    &0.399 &\phm0.205 &-0.053    &0.303 &0.391 \\
R4  &          10 &         \phm460     &\phm\phm0   & 0.155 &-0.018    &\phm0.273 &0.155 &-0.178    &0.525 &\phm0.307 &-0.163    &0.391 &0.485 \\
R5  &          25 &         \phm772     &\phm\phm0   & 0.171 &\phm0.072 &-0.250    &0.172 &-0.006    &0.598 &-0.308    &-0.148    &0.440 &0.533 \\
R6  &         200 &        2358         &\phm\phm0   & 0.188 &-0.035    &\phm0.209 &0.189 &-0.340    &0.698 &\phm0.207 &-0.313    &0.506 &0.586 \\
R1' &           1 &         \phm128     &\phm0.2     & 0.127 &\phm0.060 &-0.126    &0.124 &\phm0.034 &0.400 &-0.007    &\phm0.026 &0.321 &0.413 \\
R2' &           4 &         \phm289     & 0.09       & 0.154 &-0.024    &\phm0.036 &0.151 &-0.313    &0.520 &\phm0.017 &-0.273    &0.395 &0.494 \\
R3' &          25 &         \phm764     & 0.03       & 0.164 &\phm0.013 &-0.087    &0.165 &-0.148    &0.593 &-0.062    &-0.148    &0.438 &0.528 \\[.5mm]
\hline
\hline
   \end{array}
   $$ 
\end{table*}
%}

%\onltab{4}{
\begin{table*}
  \centering
  \caption[]{Summary of the 
  DNS results with box aspect ratio $\Gamma=4$; $\Ray=3\cdot 10^5$, $\Tay=1.6\cdot 10^7$. For normalizations see Table~\ref{tab:bres}. 
  The grid resolution is $64^2\times 256$ in all cases.}
  \vspace{-0.5cm}
  \label{tab:bresGam}
  \arraycolsep5pt
  $$
  \begin{array}{rrrrrrrrrrrrr}
    \hline
    \noalign{\smallskip}
  \vartheta & \Rey & \Co & \trxx & \trxy/10^{-2} & \trxz/10^{-2} & \tryy & \tryz/10^{-2} & \trzz & \tilde{\meanF}_x/10^{-2} & \tilde{\meanF}_y/10^{-2} & \tilde{\meanF}_z & \tilde{\meanQ} \\
    \noalign{\smallskip}
    \hline\hline\\[-3mm]
     0\degr  &  86   & 4.84   &  0.143 &  0.009     &  -0.105  &  0.144 &  -0.079     &  0.302  &   -0.036     & -0.055  & 0.273  & 0.544 \\
   15\degr  &  81   & 5.15   &  0.134 &  0.180     &  -1.828  &  0.129 &  -1.174    &  0.257  &   -3.193     & -2.494  & 0.227  & 0.440 \\
   30\degr  &  89   & 4.74   &  0.180 &  0.445     &   1.939  &  0.138 &  -1.962     &  0.311  &  -1.862      & -5.249  & 0.250  & 0.465 \\
   45\degr  & 110  & 3.81   &  0.236 &  0.986     &   6.390  &  0.197 &  -2.989    &  0.517  &    0.007     & -9.363  & 0.367  & 0.702 \\
   60\degr  & 107  & 1.69   &  0.209 &  1.401     &   1.100  &  0.206 &  -1.381    &  0.482  &   -0.885     & -8.580  & 0.314  & 0.624 \\
   75\degr  &  88   & 4.72   &  0.193 &  2.500     &   0.275  &  0.174 &   0.369     &  0.241  &    0.425     & -3.576  & 0.147  & 0.382 \\
   90\degr  &  53   & 7.82   &  0.003 & -0.011     &   -0.002 &  0.161 &  -0.111    &  0.053  &   -0.001     & -0.556  & 0.023  & 0.114 \\[.5mm]
\hline
\hline
 \end{array}
 $$ 
\end{table*}
%}
%\end{document}

%MJK AN format
%\bibliographystyle{aa}
%\bibliography{paper}

%\vspace{1cm} \noindent {\small \emph{$ $Id: paper.tex 5128 2013-05-11 16:04:41Z kapyla $ $}

\end{document}